\documentclass[twocolumn,tighten]{aastex62}
\accepted{to AJ December 6, 2018}
\received{October 10, 2018}
\submitjournal{AAS journals}
\usepackage{graphicx}
\usepackage{natbib}
\usepackage{latexsym}
\usepackage{amssymb}
\usepackage{longtable}
\usepackage{amsmath}
\usepackage{url}
\def\msun{\ifmmode {\rm\,M_\odot}\else ${\rm\,M_\odot}$\fi}
\def\Msun{\ifmmode {\rm\,\it{M_\odot}}\else ${\rm\,M_\odot}$\fi}
\def\lsun{\ifmmode {\rm\,L_\odot}\else ${\rm\,L_\odot}$\fi}
\def\Lsun{\ifmmode {\rm\,\it{L_\odot}}\else ${\rm\,L_\odot}$\fi}
\def\rsun{\ifmmode {\rm\,R_\odot}\else ${\rm\,R_\odot}$\fi}
\def\Rsun{\ifmmode {\rm\,\it{R_\odot}}\else ${\rm\,R_\odot}$\fi}
\def\Tsun{\ifmmode {\rm\,T_\odot}\else ${\rm\,T_\odot}$\fi}
\def\arcsec{\ifmmode {^{\prime\prime}}\else $^{\prime\prime}$\fi}
\def\asec{\ifmmode {^{\prime\prime}}\else $^{\prime\prime}$\fi}
\def\arcmin{\ifmmode {^{\prime}}\else $^{\prime}$\fi}
\def\amin{\ifmmode {^{\prime}}\else $^{\prime}$\fi}
\def\simlt{\mathrel{\spose{\lower 3pt\hbox{$\mathchar"218$}}
     \raise 2.0pt\hbox{$\mathchar"13C$}}}
\def\simgt{\mathrel{\spose{\lower 3pt\hbox{$\mathchar"218$}}
\     \raise 2.0pt\hbox{$\mathchar"13E$}}}

\begin{document}

\author{P. Wilson Cauley}
\affiliation{Laboratory for Atmospheric and Space Physics, University of Colorado Boulder, Boulder, CO 80303}
\affiliation{Arizona State University, School of Earth and Space Exploration, Tempe, AZ 85287}

\author{Evgenya L. Shkolnik}
\affiliation{Arizona State University, School of Earth and Space Exploration, Tempe, AZ 85287}

\author{Ilya Ilyin}
\affiliation{Leibniz-Institute for Astrophysics Potsdam (AIP), An der Sternwarte 16, 14482, Potsdam, Germany}

\author{Klaus G. Strassmeier}
\affiliation{Leibniz-Institute for Astrophysics Potsdam (AIP), An der Sternwarte 16, 14482, Potsdam, Germany}

\author{Seth Redfield}
\affiliation{Wesleyan University, Astronomy Department, Van Vleck Observatory, Middletown, CT}

\author{Adam Jensen}
\affiliation{Department of Physics and Astronomy, University of Nebraska at Kearney, Kearney, NE 68849}

\correspondingauthor{P. Wilson Cauley}
\email{pwcauley@gmail.com}

\title{Atmospheric dynamics and the variable transit of KELT-9 b\footnote{Based on data acquired with PEPSI
using the Large Binocular Telescope (LBT). The LBT is an international
collaboration among institutions in the United States, Italy, and Germany. LBT
Corporation partners are the University of Arizona on behalf of the Arizona
university system; Istituto Nazionale di Astrofisica, Italy; LBT
Beteiligungsgesellschaft, Germany, representing the Max-Planck Society, the
Leibniz-Institute for Astrophysics Potsdam (AIP), and Heidelberg University;
the Ohio State University; and the Research Corporation, on behalf of the
University of Notre Dame, University of Minnesota and University of Virginia.}}

\begin{abstract} 

We present a spectrally and temporally resolved detection of the
optical \ion{Mg}{1} triplet at 7.8$\sigma$ in the extended atmosphere of the
ultra-hot Jupiter KELT-9 b, adding to the list of detected metal species in the
hottest gas giant currently known. Constraints are placed on the density and
radial extent of the excited hydrogen envelope using simultaneous observations
of H$\alpha$ and H$\beta$ under the assumption of a spherically symmetric
atmosphere. We find that planetary rotational broadening of $v_\text{rot} =
8.2^{+0.6}_{-0.7}$ km s$^{-1}$ is necessary to reproduce the Balmer line
transmission profile shapes, where the model including rotation is strongly
preferred over the non-rotating model using a Bayesian information criterion
comparison. The time-series of both metal line and hydrogen absorption show
remarkable structure, suggesting that the atmosphere observed during this
transit is dynamic rather than static. We detect a relative emission feature
near the end of the transit which exhibits a P-Cygni-like shape, evidence of
material moving at $\approx 50-100$ km s$^{-1}$ away from the planet. We
hypothesize that the in-transit variability and subsequent P-Cygni-like
profiles are due to a flaring event that caused the atmosphere to expand,
resulting in unbound material being accelerated to high speeds by stellar
radiation pressure. Further spectroscopic transit observations will help
establish the frequency of such events.

\end{abstract}


\section{INTRODUCTION}
\label{sec:intro}

Massive gas giants in short orbital periods ($P_\text{orb} \lesssim 10$ days)
are laboratories for probing unusual planetary atmospheric conditions
and star-planet interactions. The high UV insolation levels experienced by
these planets can cause extreme atmospheric mass loss
\citep{vidal03,murray09,desetangs12,owen12,kulow14,ehrenreich15} and
high-velocity winds and zonal jets
\citep{showman08,rauscher10,rauscher14,louden15,brogi16}, while orbits near the
Alfv\'{e}n radius can lead to observable magnetic interactions between the
planet and star
\citep{cuntz00,shkolnik05,shkolnik08,lanza09,strugarek14,cauley18a} and
pre-transit signatures due to shocked stellar or planetary wind material
\citep{fossati10,vidotto10,llama11,llama13,cauley15,cauley18b}. 

The atmospheres of transiting hot planets are readily probed via transmission
spectroscopy, which can reveal the presence of individual atomic
\citep[e.g.,][]{redfield08,snellen08,jensen11,jensen12,pont13,cauley15,wilson15,wyttenbach15,wyttenbach17,casasayas17,casasayas18,spake18,jensen18}
and molecular species
\citep[e.g.,][]{knutson07,snellen10,deming13,kreidberg15,brogi16,sing16},
measure the dynamics of planetary rotation and winds
\citep{bourrier15,louden15,brogi16}, and detect hazes and cloud layers
\citep{knutson14,mallonn16,kreidberg18,chen18}. Optical transmission spectra
generally target strong atomic lines, such as \ion{Na}{1} D, \ion{K}{1} 7698
\AA, and H$\alpha$, which probe the thermosphere at pressures of $\approx
1\mu$bar \citep{heng15,huang17}. Recently, \ion{He}{1} 10830 \AA\ was detected
in the extended atmospheres of WASP-107 b and HAT-P-11 b, demonstrating the
line's potential as a mass loss diagnostic for hot planets
\citep{spake18,mansfield18,oklopcic18}. 

The hottest gas giant discovered so far is KELT-9 b \citep{gaudi17},
which orbits at a distance of $0.034$ AU from its A0V/B9V host star
($T_\text{eff} = 10170\pm450$ K) with a dayside equilibrium temperature of
$T_\text{eq} \approx 4600$ K. \citet{kitzmann18} showed that KELT-9 b's high
temperature, which should dissociate most refractory molecules into their
constituent atoms, could produce a suite of observable \ion{Fe}{1} and
\ion{Fe}{2} lines in the optical. This prediction was validated by
\citet{hoeijmakers18} who detected a strong cross-correlation signal in optical
\ion{Fe}{1}, \ion{Fe}{2}, and \ion{Ti}{2} lines. \citet{yan18}, via absorption
at H$\alpha$, detected the presence of an optically thick layer of excited
hydrogen around KELT-9 b and estimated the planetary mass loss rate to be $\sim
10^{12}$ g s$^{-1}$. 

The metal detections from \citet{hoeijmakers18} were made by cross-correlating
a large number of theoretical iron line absorption cross sections with
in-transit spectra obtained with HARPS-N on the 3.58-meter Telescopio Nazionale
Galileo (TNG). The H$\alpha$ observations of \citet{yan18} were also acquired
with a 3.5-meter telescope (the 3.5-meter at Calar Alto Observatory). While the
cross correlation technique is powerful when a large number of lines are
present in the spectrum, the detection of weak transmission spectrum absorption
of individual spectral lines during a single transit, as opposed to the strong
($\approx 1\%$) signal seen in H$\alpha$ \citep{yan18}, requires the use of
10-meter class telescopes combined with efficient high-resolution
spectrographs. 

Here we present the first exoplanet atmosphere detection of the optical
\ion{Mg}{1} triplet around KELT-9 b using the PEPSI spectrograph
\citep{strassmeier15} on the Large Binocular Telescope (LBT; $2\times8.4$ m).
Magnesium absorption can be used to estimate exoplanet mass loss rates
\citep{bourrier15} and is an important coolant in hot planet atmospheres
\citep{huang17}. We also confirm the H$\alpha$ measurement of \citet{yan18} and
provide additional constraints on the extended hydrogen atmosphere using
simultaneous observations of H$\beta$.  Atmospheric models of the \ion{Mg}{1}
and Balmer line absorption favor non-zero planetary rotation velocities.
Finally, we observe significant in-transit variability in all of the planetary
absorption lines and large blue-shifted velocities in the transmission profiles
near the end of the transit. We hypothesize that a stellar flare event is
responsible for the variability and an increase in the planet's mass loss rate,
where the expanding material is accelerated to high velocities by stellar
radiation pressure.

\section{OBSERVATIONS AND DATA REDUCTION}
\label{sec:obs}

We observed the transit of KELT-9 b between 04:07--11:20\,UT on July 3, 2018
with the Large Binocular Telescope in Arizona and its high-resolution
\'echelle spectrograph PEPSI \citep{strassmeier15}. PEPSI is a white-pupil
\'echelle spectrograph with two arms (blue and red optimized) and is equipped
with six cross dispersers (CD\,I to VI) for full optical wavelength coverage.
Sky fibers for simultaneous sky and target exposures are available but were not
used for the present observations.  In use for the data in this paper was the
new image slicer block \#3. Its $R\approx50,000$-mode employs two three-slice
waveguide image slicers and a pair of 300-$\mu$m fibers with a projected sky
aperture of 2.3\arcsec . Seeing for the night was $\approx 1''$. Two
10.3k$\times$10.3k STA1600LN CCDs with 9-$\mu$m pixels recorded a total of 34
\'echelle orders in CD III and V together. One spectral resolution element
corresponds to 0.10\,\AA\ in the blue and 0.14\,\AA\ in the red ($\approx$6\,km
s$^{-1}$) and is sampled with 12.4 pixels. The dispersion changes from 8~m\AA\
pixel$^{-1}$ at 4000\,\AA\ to 18~m\AA\ pixel$^{-1}$ at 9000\,\AA\ and results
in an unequally spaced pixel-step size.

PEPSI was used in its $R \approx 50,000$ mode and with cross dispersers (CD)
III (blue arm) and V (red arm) simultaneously. The wavelength coverage was
4750--5430\,\AA\ in the blue arm and 6230--7430\,\AA\ in the red arm. The
spectra were collected with a constant signal-to-noise of 210 in the continuum
controlled by a photon counter. This results in slightly different exposure
times for each CD and produced 82 blue arm spectra and 73 red arm spectra, with
exposure times between 220\,s and 300\,s depending on air mass. 

All data were reduced with the Spectroscopic Data System for PEPSI (SDS4PEPSI)
which is a generic software package written in C++ under a Linux environment.
The standard reduction steps include bias overscan detection and subtraction,
scattered light surface extraction from the inter-order space and subsequent
subtraction, definition of \'echelle orders, optimal extraction of spectral
orders, wavelength calibration, and a self-consistent continuum fit to the full
2D image of extracted orders. Wavelength calibration was based on a standard
Thorium-Argon (Th-Ar) hollow-cathode lamp. The spectrograph is located in a
pressure-controlled chamber at a constant temperature and humidity to keep the
refractive index of the air inside constant over a long-term period, which
provides radial velocity stability of about 5\,m s$^{-1}$.  The ThAr spectra
were taken just before observations started and no significant instabilities in
radial velocity of the spectra were detected as a function of time. We employ a
``super master'' flat fielding procedure to correct for the spatial
pixel-to-pixel noise on the CCD.  The super-master flat is a polynomial fit to
the amplitude of the spatial noise versus flux in each pixel of the master
flat, which is an average of 5,000 flat exposures. This procedure is specific
to the STA1600LN CCD. The individual science images are first flat fielded and
then optimally extracted. More details on PEPSI data reduction can be found in
\citet{strassmeier18}. The extracted spectra are corrected for the Earth's
barycentric motion at the time of the observation.

\begin{figure*}[htbp]
   \centering
   \includegraphics[scale=.70,clip,trim=5mm 5mm 10mm 20mm,angle=0]{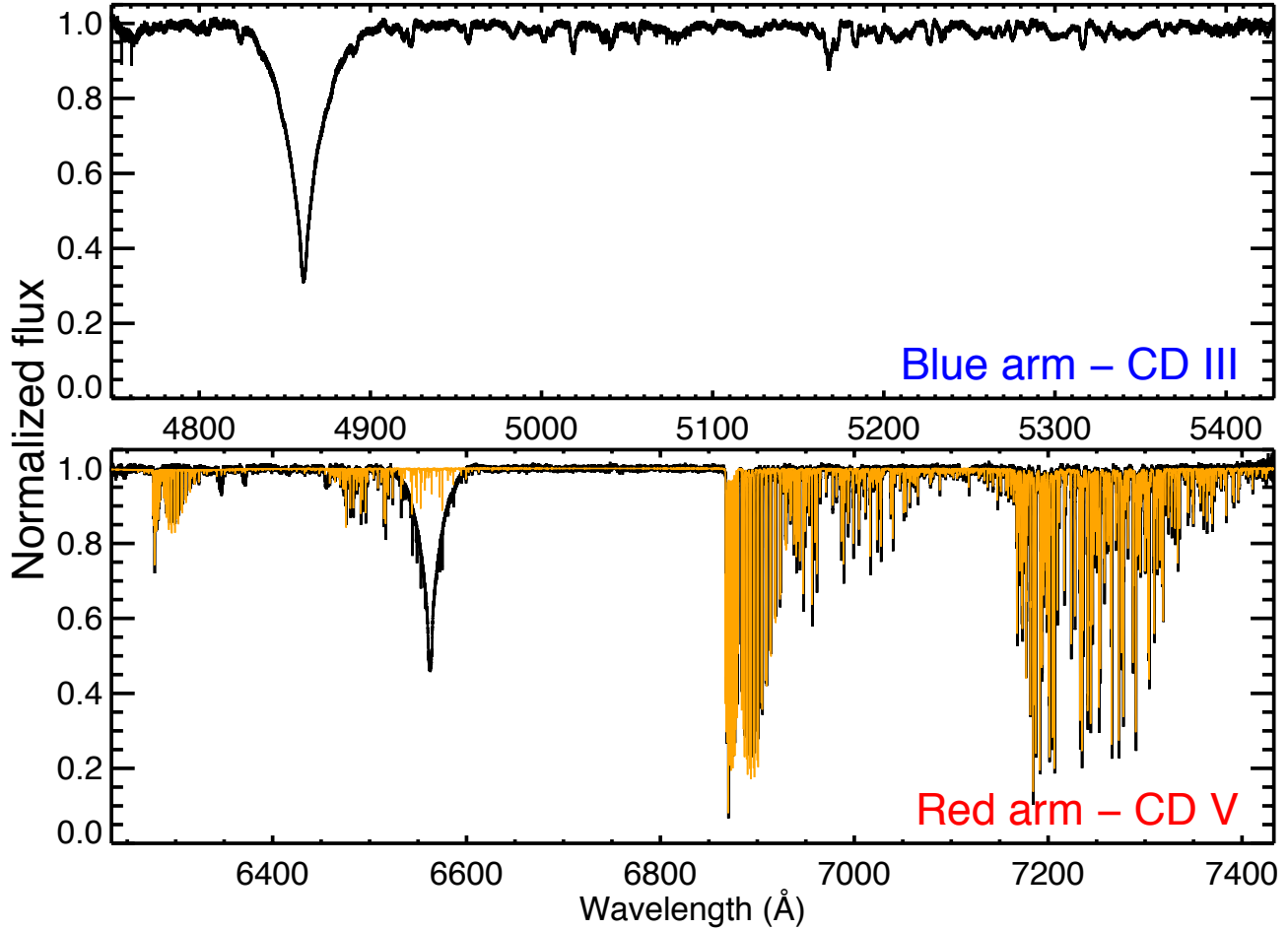}

   \figcaption{Individual KELT-9 spectrum (black) showing the full wavelength
range of a PEPSI exposure using CDs III and V. The orange spectrum is the
nominal telluric model, found with \texttt{MOLECFIT}, for the red arm.
Note that the telluric spectrum has not been scaled to precisely match the
telluric absorption lines. \label{fig:full}}

\end{figure*}

\subsection{Telluric line removal}
\label{sec:tell}

Telluric line contamination is present throughout the red arm spectrum (see
\autoref{fig:full}).  In order to remove the telluric lines, we used the
software package \texttt{MOLECFIT} \citep{kausch15,smette15}. We fit regions of
the average KELT-9 red arm spectrum that did not contain photospheric
absorption. This mainly avoided the spectrum surrounding H$\alpha$ at 6562.79
\AA. The nominal telluric model produced by \texttt{MOLECFIT} is shown in
orange in \autoref{fig:full}. We then fit the telluric absorption for
individual exposures by scaling and shifting the nominal telluric model. We
optimized the removal to focus on the telluric lines near H$\alpha$. We note
that by doing this the strong telluric lines in other regions, e.g., the O$_2$
and H$_2$O band near 7200 \AA, are not as cleanly removed. This is the
result of line blending and imperfect model profile shapes, which exacerbates
the residuals for strong lines. However, the poor correction in the strong
lines does not affect the detection of transmission signatures in regions of
weak or non-existant telluric absorption.  The unblended weak telluric lines
near H$\alpha$ are removed down to the photon noise level for individual
spectra. 

\subsection{Center-to-limb variations}

Center-to-limb variations, or CLVs, in spectral lines are the result of
wavelength-dependent limb darkening across the stellar disk. CLVs can produce
spurious transmission spectrum signals if not properly taken into account
\citep[e.g.,][]{yan15,czesla15,khalafinejad17,yan17,cauley17a}. In order to
remove the CLVs from our transmission spectra, we use the \texttt{Spectroscopy
Made Easy} (\texttt{SME}) package \citep{valenti96,piskunov17} to generate
synthetic spectra as a function of $\mu = \cos\theta$. We adopt the KELT-9
stellar parameters from \citet{gaudi17} for the model spectra (\autoref{tab:pars}). The spectra are
generated at 25 different $\mu$-values and interpolated onto the final stellar
$\mu$-grid used in the transit simulations.

\begin{deluxetable}{lccc}
\tablecaption{KELT-9 system parameters\label{tab:pars}}
\tablehead{\colhead{Parameter}&\colhead{Symbol}&\colhead{Units}&\colhead{Value}}
\colnumbers
\tabletypesize{\scriptsize}
\startdata
Stellar mass & $M_\star$ & $M_\Sun$ & $2.52^{+0.25}_{-0.20}$ \\
Stellar radius & $R_\star$ & $R_\Sun$ & $2.362^{+0.075}_{-0.063}$ \\
Effective temperature & $T_\text{eff}$ & K & $10170 \pm 450$ \\
Metallicity & [Fe/H] & \nodata & $-0.03 \pm 0.20$ \\
Stellar rotational velocity & $v$sin$i$ & km s$^{-1}$ & $111.4 \pm 1.3$ \\
Spin-orbit alignment angle & $\lambda$ & degrees & $-84.8\pm1.4$ \\ 
Orbital period & $P_\text{orb}$ & days & $1.4811235 \pm 0.0000011$ \\
Semi-major axis & $a$ & AU & $0.03462^{+0.00110}_{-0.00093}$ \\
Planetary mass & $M_\text{p}$ & $M_\text{J}$ & $2.88 \pm 0.84$ \\
Planetary radius & $R_\text{p}$ & $R_\text{J}$ & $1.891^{+0.061}_{-0.053}$ \\
Orbital velocity$^\dagger$ & $v_\text{orb}$ & km s$^{-1}$ & $254.3 \pm 8.1$ \\
\enddata
\tablenotetext{\dagger}{Calculated from $P_\text{orb}$ and $a$ assuming zero
eccentricity.}
\end{deluxetable}

The in-transit CLV contributions are calculated by simulating the transit of
KELT-9 b at the same times as our observations, taking into account the
misaligned orbit of the planet and rotational broadening due to the $v$sin$i$
of the star. By taking into account stellar rotation and spin-orbit
angle, we are also correcting for velocity shifts in the observed spectrum due
to the Rossiter-McLaughlin (RM) effect. The planetary and orbital parameters
are taken from \citet{gaudi17} (\autoref{tab:pars}). We generate a spatial grid with resolution
$0.01 R_\star \times 0.01 R_\star$ with the appropriate CLV- and RM-modified
stellar spectrum assigned to each grid cell. We then integrate the in-transit
stellar spectrum at the wavelengths of interest. The in-transit spectrum is
then divided by the integrated out-of-transit spectrum.  The result is the
in-transit profile surrounding the desired spectral line. The in-transit
profile can then be divided out of the observed transmission spectrum, leaving
behind any atmospheric contribution from the planet
\citep[e.g.,][]{khalafinejad17,yan18}. 

\section{RESULTS}
\label{sec:results}

We identified absorption in the planetary atmosphere by creating an average
transmission spectrum, which we denote as $S_\text{T} =
S_\text{in}/S_\text{out}$, for each arm and automatically searching for
significant deviations in the spectrum. The out-of-transit spectra ($N=37$ for
the blue arm and $N=29$ for the red arm) were averaged to create a master
$S_\text{out}$ spectrum. The average $S_\text{T}$ spectrum is the mean of all
$S_\text{T}$ obtained between the second and third transit contact points,
where the planet does not intersect the stellar limb.  We shifted the
individual $S_\text{T}$ spectra into the rest frame of the planet, eliminating
the orbital velocity signature. The RV of the star is also removed.  We
performed the line search using Gaussian fits across the entire observed
wavelength range and flagging any features with depths $\geq 1.5 \sigma$ below
the local continuum and FWHM values of $\geq 0.25$ \AA. We then manually
checked the flagged wavelengths confirm the presence of an absorption line;
spurious signatures, e.g., artifacts from joined orders with overlapping
wavelength regions, were discarded. 

Absorption values for lines with detections at $\geq 3\sigma$ are given in
\autoref{tab:abslines}, along with line depths, central velocities, and
FWHM values derived by fitting a Gaussian using a Markov Chain Monte Carlo
(MCMC) procedure based on the algorithm by \citet{goodman10} \citep[see
also][]{foreman12}. The absorption, which is given in equivalent angstroms,
is calculated by integrating the transmission spectrum within $\pm50$ km
s$^{-1}$ of the line's central air wavelength for the Balmer lines and
$\pm30$ km s$^{-1}$ for the narrower metal lines. The uncertainty in the
average equivalent width $W_\lambda$ values is calculated by measuring the
standard deviation of $S_T$ outside of the the absorption integration
region but within $\pm 200$ km s$^{-1}$ and assigning this value as the
error for each point in $S_T$.  Detected lines with overlapping $\pm 200$
km s$^{-1}$ segments are ignored when calculating the absorption in one of
the lines. The individual errors are then summed in quadrature to produce
the final uncertainty given in column 4 of \autoref{tab:abslines}. The
average $S_\text{T}$ profiles for the lines in \autoref{tab:abslines} are
shown in \autoref{fig:tspecs_all} ordered by atomic weight and sub-ordered
by wavelength. The uncertainties on the Gaussian line parameters are the
68\% values from the marginalized posterior distributions.

\begin{deluxetable*}{lcccccc}
\tablewidth{1.99\textwidth}
\tablecaption{Average in-transit absorption and line parameters\label{tab:abslines}}
\tablehead{\colhead{}&\colhead{$\lambda_0$}&\colhead{$W_\lambda$}&\colhead{$\sigma_\text{W}$}&\colhead{Depth}&\colhead{$v_\text{cen}$}&\colhead{FWHM}\\
\colhead{Species}&\colhead{(\AA)}&\colhead{(m\AA)}&\colhead{(m\AA)}&\colhead{(\%)}&\colhead{(km s$^{-1}$)}&\colhead{(km s$^{-1}$)}}
\colnumbers
\startdata
\ion{Ti}{2} & 4805.09 & 0.60 & 0.15 & $0.102^{+0.019}_{-0.016}$ & $-5.27^{+2.61}_{-2.67}$ & $30.5^{+15.1}_{-5.0}$ \\
\ion{Fe}{1} & 4824.17 & 0.58 & 0.16 & $0.134^{+0.014}_{-0.015}$ & $-3.20^{+1.78}_{-1.56}$ & $27.0^{+4.6}_{-3.2}$ \\
H$\beta$ & 4861.35 & 6.45 & 0.27 & $0.783^{+0.009}_{-0.010}$ & $-0.97^{+0.35}_{-0.33}$ & $49.8^{+0.7}_{-0.7}$ \\
\ion{Fe}{2} & 4923.92 & 1.86 & 0.14 & $0.414^{+0.017}_{-0.016}$ & $1.76^{+0.45}_{-0.54}$ & $24.7^{+1.5}_{-1.2}$ \\
\ion{Fe}{1}$^*$ & 4957.60 & 0.89 & 0.13 & $0.121^{+0.012}_{-0.013}$ & $-8.62^{+3.94}_{-7.84}$ & $56.7^{+33.8}_{-13.1}$ \\
\ion{Fe}{2} & 5018.44 & 2.28 & 0.14 & $0.460^{+0.017}_{-0.015}$ & $1.85^{+0.45}_{-0.64}$ & $26.9^{+1.4}_{-1.1}$ \\
\ion{Mg}{1} & 5167.32 & 0.99 & 0.13 & $0.157^{+0.013}_{-0.014}$ & $4.52^{+1.94}_{-1.85}$ & $38.9^{+7.7}_{-4.7}$ \\
\ion{Fe}{2} & 5169.03 & 2.40 & 0.12 & $0.475^{+0.015}_{-0.014}$ & $1.68^{+0.40}_{-0.44}$ & $26.9^{+1.2}_{-1.0}$ \\
\ion{Mg}{1} & 5172.68 & 1.01 & 0.13 & $0.172^{+0.016}_{-0.015}$ & $-1.33^{+1.25}_{-1.18}$ & $30.3^{+4.8}_{-3.2}$ \\
\ion{Mg}{1} & 5183.60 & 0.99 & 0.12 & $0.157^{+0.016}_{-0.017}$ & $-0.98^{+1.30}_{-1.44}$ & $31.8^{+6.9}_{-4.1}$ \\
\ion{Ti}{2} & 5188.69 & 0.61 & 0.12 & $0.117^{+0.019}_{-0.016}$ & $-1.30^{+1.85}_{-1.77}$ & $29.6^{+11.4}_{-4.5}$ \\
\ion{Fe}{2} & 5197.57 & 0.98 & 0.13 & $0.184^{+0.015}_{-0.014}$ & $1.43^{+1.08}_{-1.13}$ & $28.9^{+3.8}_{-2.7}$ \\
\ion{Ti}{2} & 5226.54 & 0.88 & 0.13 & $0.142^{+0.014}_{-0.014}$ & $2.89^{+1.64}_{-1.63}$ & $32.8^{+6.3}_{-4.0}$ \\
\ion{Fe}{2} & 5234.62 & 1.09 & 0.13 & $0.173^{+0.012}_{-0.014}$ & $0.92^{+1.23}_{-1.21}$ & $34.7^{+4.7}_{-3.5}$ \\
\ion{Fe}{2} & 5275.99 & 1.08 & 0.15 & $0.268^{+0.017}_{-0.015}$ & $1.18^{+0.61}_{-0.66}$ & $20.9^{+1.8}_{-1.3}$ \\
\ion{Fe}{2} & 5316.61 & 1.72 & 0.14 & $0.317^{+0.015}_{-0.014}$ & $2.54^{+0.60}_{-0.72}$ & $28.0^{+2.1}_{-1.7}$ \\
\ion{Fe}{2} & 5362.86 & 0.69 & 0.14 & $0.154^{+0.016}_{-0.015}$ & $-1.02^{+1.08}_{-1.26}$ & $23.7^{+3.6}_{-2.3}$ \\
\ion{Fe}{2} & 6456.38 & 1.34 & 0.13 & $0.181^{+0.023}_{-0.017}$ & $3.00^{+1.17}_{-1.46}$ & $24.7^{+5.0}_{-2.7}$ \\
H$\alpha$ & 6562.79 & 13.67 & 0.208 & $1.103^{+0.010}_{-0.009}$ & $2.67^{+0.25}_{-0.28}$ & $55.1^{+0.6}_{-0.6}$ \\
\enddata
\tablenotetext{*}{The large negative value of $v_\text{cen}$ and its large error bars are due to a blend with
\ion{Fe}{1} 4957.298 \AA.}
\end{deluxetable*}

\subsection{Hydrogen absorption}
\label{sec:hydro}

We confirm the H$\alpha$ absorption detected by \citet{yan18} and report
significant absorption in H$\beta$. The H$\alpha$ and H$\beta$ line profiles are
shown in the first row of \autoref{fig:tspecs_all}.  The depth
($1.103^{+0.010}_{-0.009}$\%) and line width (FWHM $= 55.1^{+0.7}_{-0.6}$ km
s$^{-1}$) of H$\alpha$ is very similar to the \citet{yan18} measurements
($1.15\pm0.05$\% and 51.2$^{+2.7}_{-2.5}$ km s$^{-1}$, respectively) suggesting
that the excited hydrogen layer is non-variable from epoch to epoch. H$\beta$
has a line depth of $0.783^{+0.009}_{-0.010}$\% and FWHM$=49.8^{+0.7}_{-0.7}$ km
s$^{-1}$. The large H$\beta$ line depth verifies that the hydrogen layer
detected by \citet{yan18} is optically thick. 

\subsection{Metal line absorption}
\label{sec:metals}

\begin{figure*}[htbp]
   \centering
   \includegraphics[scale=.95,clip,trim=15mm 40mm 10mm 1mm,angle=0]{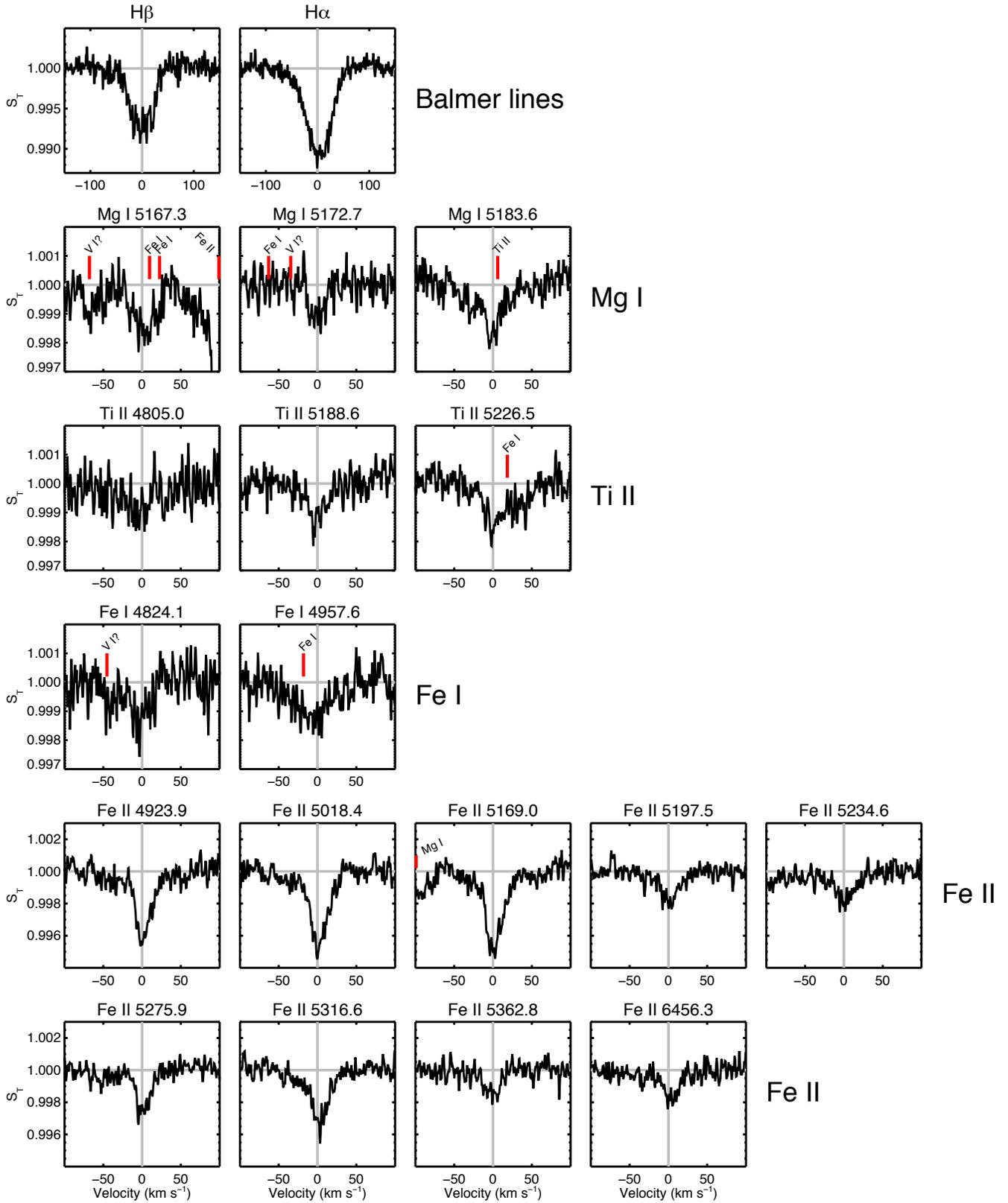}

   \figcaption{Average $S_\text{T}$ profiles for the lines with absorption
detected at $\geq 3\sigma$ listed in order of atomic weight and from
red to blue wavelengths within atomic species. Possible contaminating atomic
transitions are marked with vertical red lines. All x-axis values are velocities
in km s$^{-1}$. All lines of a particular atomic species are on the same
vertical scale. \label{fig:tspecs_all}}

\end{figure*}

\begin{figure}[htbp]
   \centering
   \includegraphics[scale=.56,clip,trim=65mm 40mm 50mm 65mm,angle=0]{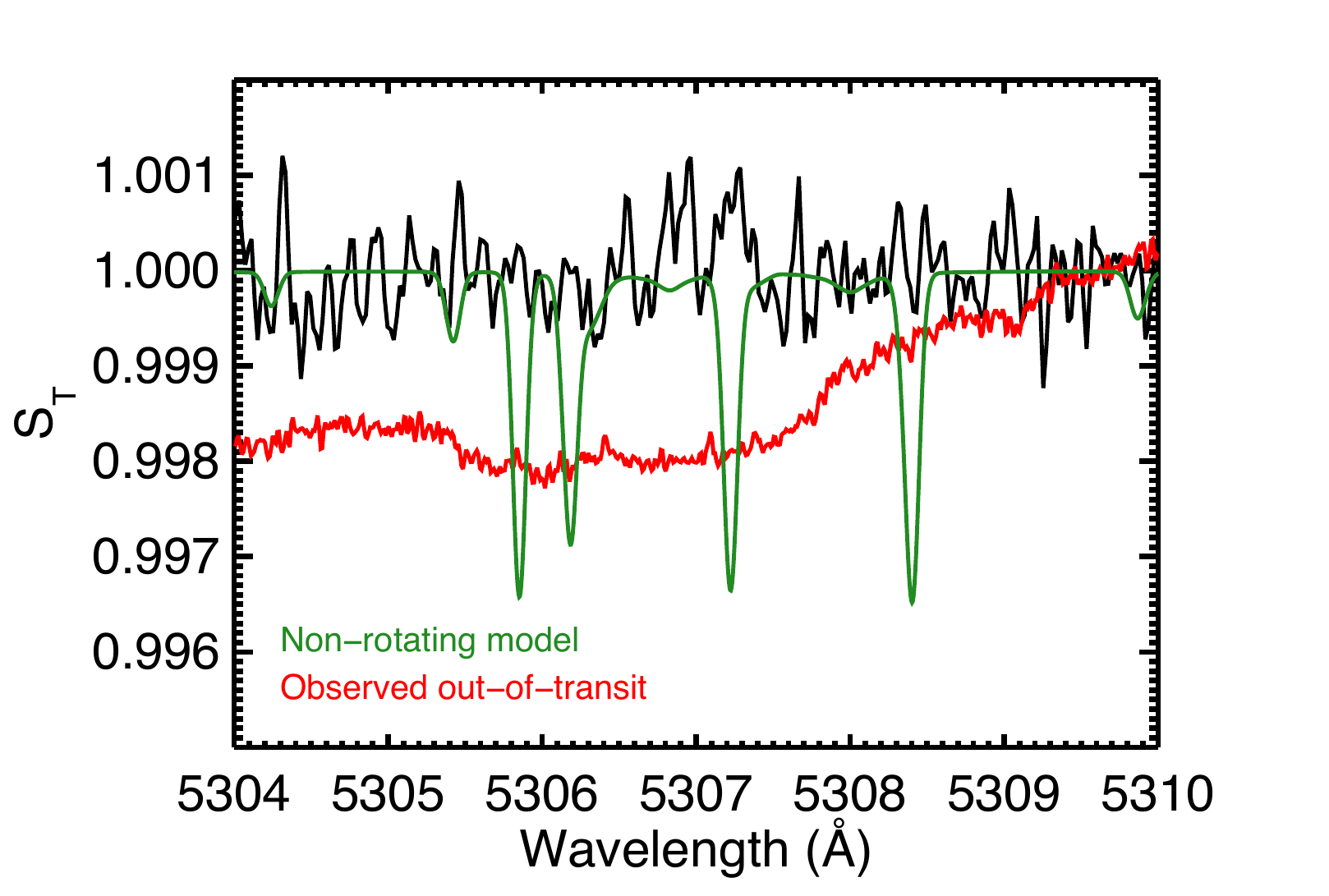}

   \figcaption{Average in-transit control region transmission spectrum. The
green line shows the non-rotating stellar model and the presence of the strong
lines \ion{Cr}{2} 5305.85 \AA, \ion{Fe}{2} 5306.18 \AA, \ion{Ca}{2} 5307.22
\AA, and \ion{Cr}{2} 5308.41 \AA\ in the stellar spectrum. The observed
rotationally broadened KELT-9 out-of-transit spectrum is shown in red.  There
is no absorption present in the control region, demonstrating that the measured
absorption in the metal lines is not an artifact. \label{fig:control}}

\end{figure}

In addition to the \ion{Fe}{1}, \ion{Fe}{2}, and \ion{Ti}{2} signatures
detected by \citet{hoeijmakers18} we report the presence of absorption in the
\ion{Mg}{1} 5167.3, 5172.7, 5183.6 \AA\ triplet. The average $S_T$ spectra for
all metal lines with absorption detected at $\geq 3\sigma$ are shown in
\autoref{fig:tspecs_all}. The relative absorption depths of
\ion{Fe}{1}, \ion{Fe}{2}, and \ion{Ti}{2} are similar to the relative S/N of
the cross-correlation functions from \citet{hoeijmakers18}. We also analyze a
control region of the spectrum where we do not expect to see planetary
absorption. This region is shown in \autoref{fig:control} and consists of
\ion{Cr}{2} and \ion{Fe}{2} lines with ground state energy levels of $E_1
\approx 4-13$ eV, too high to be significantly populated in the atmosphere of
KELT-9 b. Note that the abundance of hydrogen is what allows the $n=2$ state,
which has an excitation energy of 10.2 eV, to be populated via radiative
excitation \citep{huang17} whereas less abundant metal species simply do not
have sufficient number densities to produce observable populations of highly
excited states.
 
The \ion{Mg}{1} triplet shows absorption in each of the three lines. However,
there is considerable contamination from nearby \ion{Fe}{1}, \ion{Fe}{2}, and
\ion{Ti}{2} lines (red lines in \autoref{fig:tspecs_all}).  The triplet member
at 5172.7 \AA\ is relatively free of contaminants near the line core.  The
absorption in this line alone is significant at the level of $7.8\sigma$.  The
likelihood of the absorption being due to \ion{Mg}{1} is strengthened by the
similar absorption in the other triplet members.  

\subsection{Extended atmosphere models and planetary rotation}
\label{sec:atmomods}

In order to place constraints on the density and structure of the extended
atmosphere, we simulated the Balmer line and \ion{Mg}{1} absorption using
spherically symmetric atmospheres. We take the density in the atmosphere to be
uniform. The uniform density case for hydrogen is consistent with the results
of \citet{christie13} and \citet{huang17}, who find that the $n=2$ hydrogen
density is approximately constant across three orders of magnitude in pressure
for the case of HD 189733 b's atmosphere. For the \ion{Mg}{1} triplet, we do
not have strong constraints on the density due to the lack of additional
\ion{Mg}{1} transitions. As we will show, our \ion{Mg}{1} measurement is
consistent with an optically thin atmosphere and the density and radial extent
of the atmosphere are not well constrained. Thus assuming a uniform density is
a good first approximation.

The models consist of a 3-D grid, with spatial resolution $0.01 R_\star \times
0.01 R_\star \times 0.01 R_\star$, filled with material of number density $n$
between $R_\text{p}$ and the radial extent of the atmosphere $r$.  After the
3-D density grid is populated, the column density is calculated at each point
intersecting the stellar disk. The 2-D column density grid is then used to
extinct the stellar spectrum at the corresponding disk location. The extincted
line profiles are then summed and the ratio is taken with the out-of-transit
spectrum to create the model $S_\text{T}$ for the mean observed transit time.
A Gaussian is assumed for the intrinsic line profile shape. 

In addition to thermal broadening, which we take to be the intrinsic full-width
at half maximum of the Gaussian $v_\text{t}$, we also consider the effects of
atmospheric rotational broadening. KELT-9 b's orbital period implies a tidally
locked equatorial rotational velocity of 6.6 km s$^{-1}$. This is
non-negligible when compared with the line profile widths and should contribute
to the shapes of the lines. The planetary atmosphere is taken to be rigidly
rotating with equatorial velocity $v_\text{rot}$ and the planet's spin axis is
assumed to be perpendicular to the orbital plane. Each absorption line profile
in the planet's atmosphere is shifted by the appropriate velocity before being
applied to the stellar spectrum. 


We fit the Balmer and \ion{Mg}{1} profiles with the same MCMC procedure used to
derive the Gaussian line parameters in \autoref{sec:results}. Due to
noticeable variability in the transit absorption (see
\autoref{sec:timeseries}), we only fit the Balmer line transmission spectra
between $-0.07\, \text{days} < t-t_\text{mid} < -0.03 \, \text{days}$. This
portion of the transit is well described by a spherically symmetric atmosphere
while the transit variability suggests a sharp departure from this model. The
\ion{Mg}{1} transmission spectra do not have sufficient S/N to exclude any
in-transit measurements. Thus with the limitations of the spherically symmetric
model in mind, we fit the full \ion{Mg}{1} in-transit profile. For the Balmer
lines, we also fit a baseline model that does not include rotation.  This
allows a statistical comparison of the rotation versus non-rotation cases using
the Bayesian information criterion for each model. 

A single model line profile is calculated at the average transit time of the
observations. For the Balmer lines, the chain consisted of 100 walkers
that were run for 600 steps. We then discarded the first 100 steps as burn-in
for a total of 50,000 independent samplings of the posterior distribution.
The \ion{Mg}{1} parameters are less constrained and thus require a
longer burn-in phase. The \ion{Mg}{1} chains are run for 3000 steps and we
discard the first 1000 steps. The free parameters in the model are the number
density $n$, radial extent $R_\text{p}+r$, thermal broadening $v_\text{t}$, and
rotational velocity $v_\text{rot}$. Non-restrictive uniform priors were assumed
for all parameters.  We simultaneously fit both the H$\alpha$ and H$\beta$ line
profiles. For \ion{Mg}{1}, all of the triplet members have the same oscillator
strength and they all show similar $S_\text{T}$ profiles, i.e., no information
is gained by fitting all three simultaneously. For this reason, we only fit the
\ion{Mg}{1} 5172.7 \AA\ line since it is relatively free of nearby
contaminating lines.

\begin{figure*}[htbp]
   \centering
   \includegraphics[scale=.85,clip,trim=15mm 15mm 30mm 15mm,angle=0]{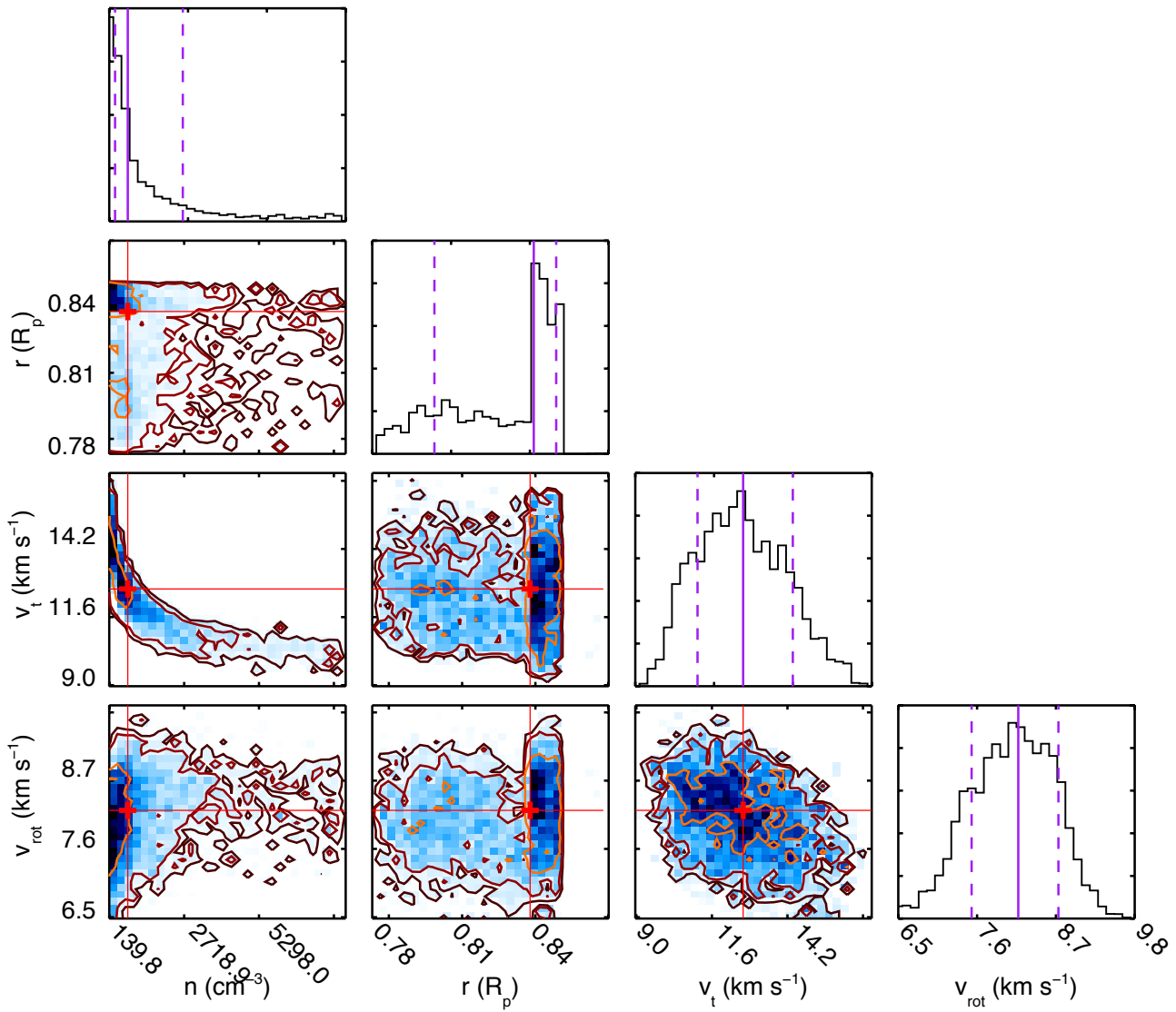}

   \figcaption{Posterior distributions and marginalized histograms for the
atmosphere fits to the Balmer lines. The median values are marked on the
posterior distributions with a red square and on the marginalized histograms
with a purple line. The 68\% confidence intervals for each parmeter are marked
with dashed purple lines. The 68\%, 95\%, and 99\% contours are also overplotted.
\label{fig:balmcorn}}

\end{figure*}

\begin{figure*}[htbp]
   \centering
   \includegraphics[scale=.85,clip,trim=15mm 15mm 30mm 15mm,angle=0]{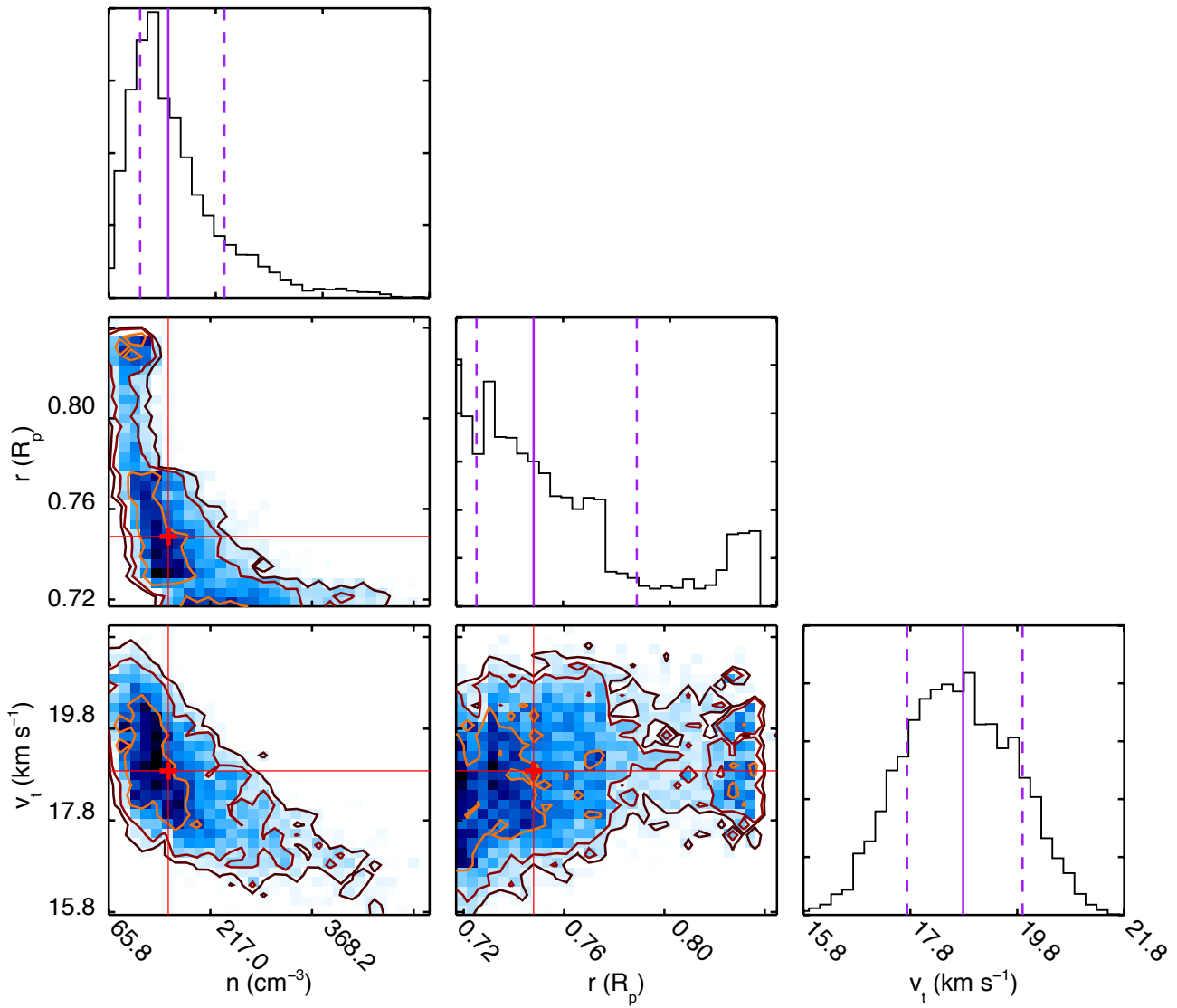}

   \figcaption{Posterior distributions and marginalized histograms for the
atmosphere fits to the Balmer lines in the case of no rotation. The
format is the same as \autoref{fig:balmcorn}
\label{fig:balmcorn_norot}}

\end{figure*}

\begin{figure*}[htbp]
   \centering
   \includegraphics[scale=.85,clip,trim=15mm 15mm 30mm 15mm,angle=0]{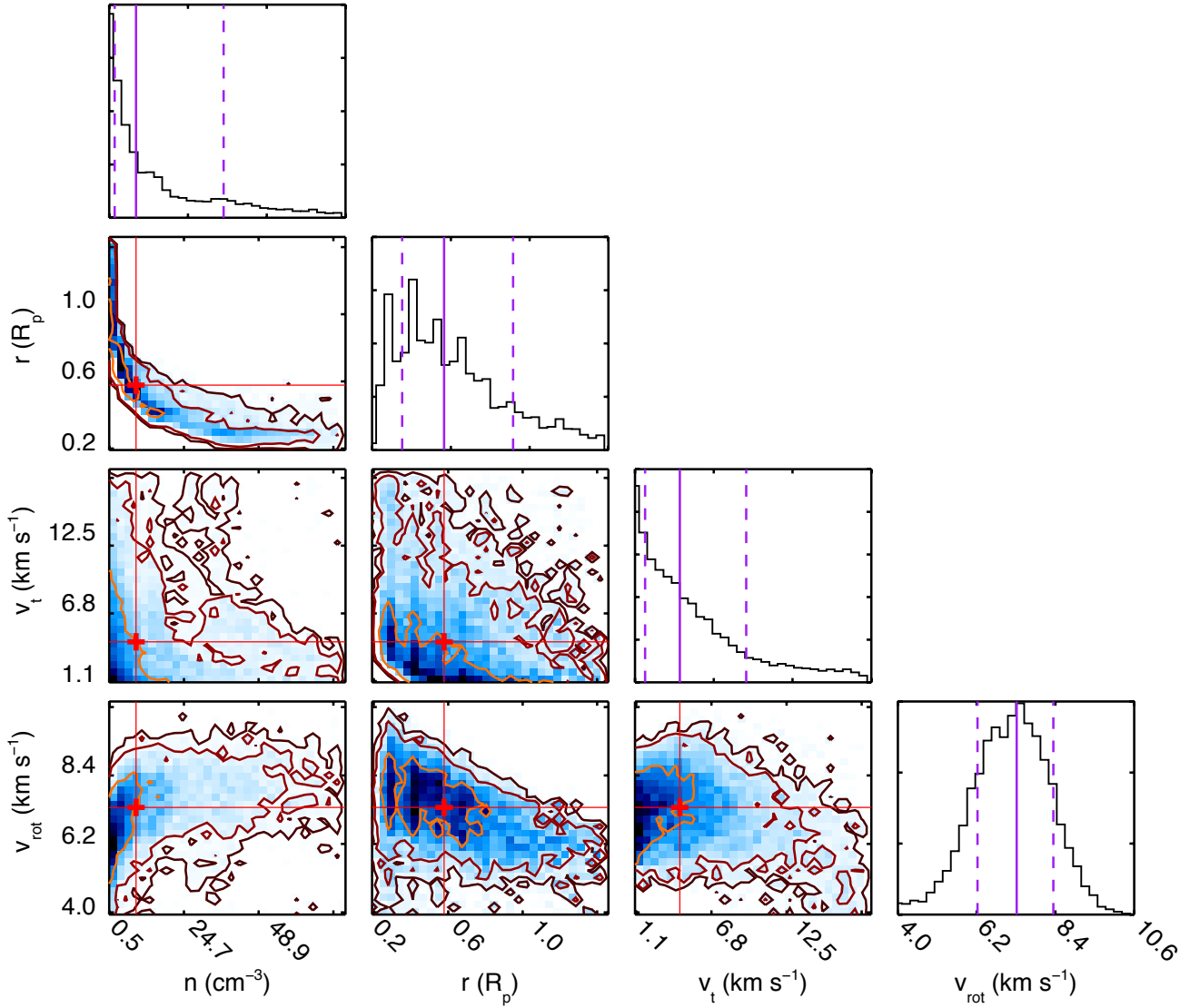}

   \figcaption{Same format as \autoref{fig:balmcorn} except for the \ion{Mg}{1}
lines. The density $n$ and radial atmospheric extent $r$ are directly
correlated since the \ion{Mg}{1} lines are optically thin, which allows a broad
range of values for $n$ and $r$. Additional \ion{Mg}{1} transitions are needed
to better constrain the atmosphere probed by the \ion{Mg}{1} optical triplet.
\label{fig:mgicorn}}

\end{figure*}

\begin{figure*}[ht!]
   \centering
   \includegraphics[scale=.7,clip,trim=0mm 10mm 20mm 30mm,angle=0]{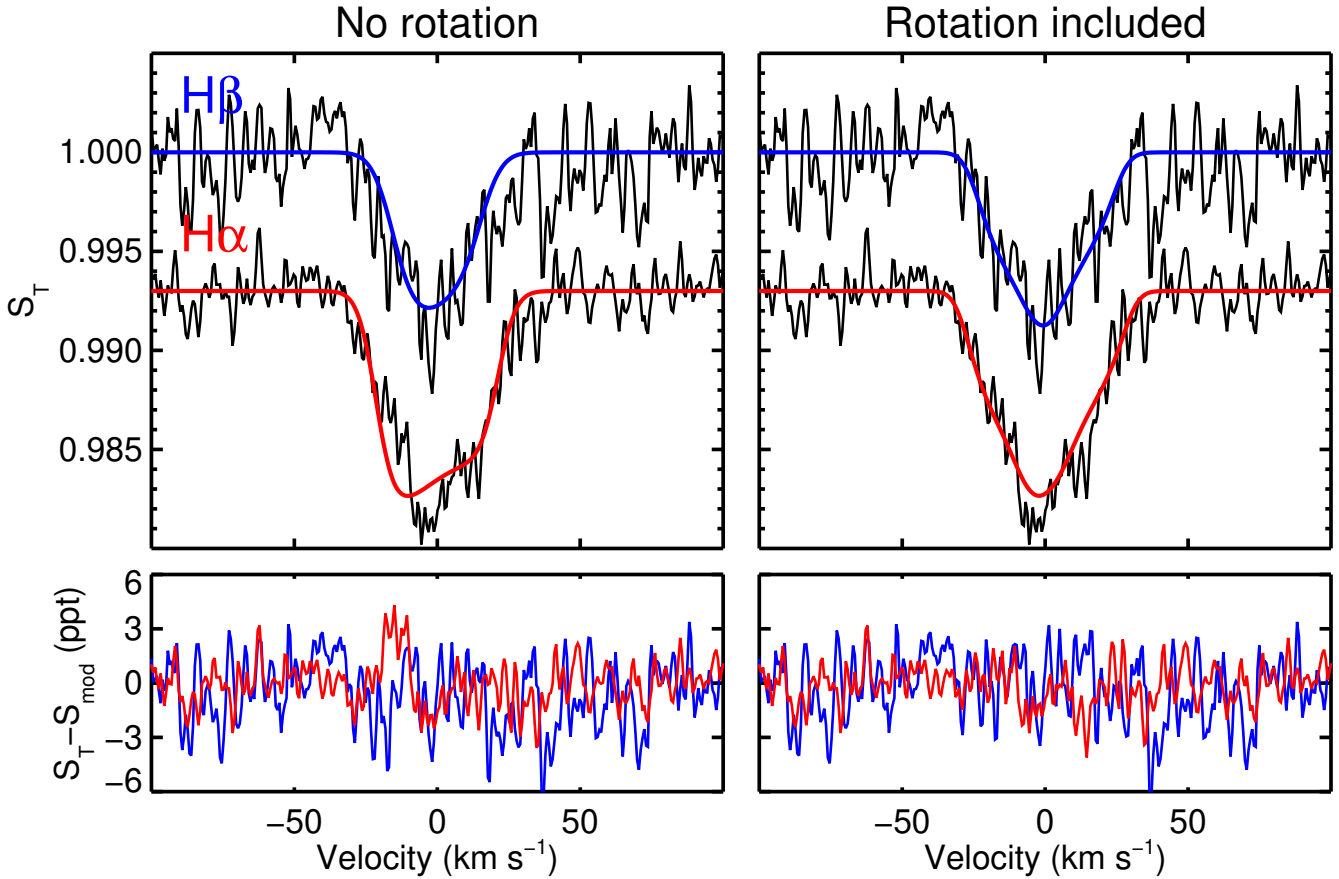}

   \figcaption{Best-fit Balmer line models for the case of no rotation (left
panels) and including rotation (right panels). The fit residuals are shown in
the bottom panels.  The H$\alpha$ data is shown in black and the H$\beta$ data
is shown in dark gray.  The H$\alpha$ models are plotted as red lines; the
H$\beta$ models as blue lines.  The rotation model provides better fits to
Balmer lines and has a substantially lower BIC value, indicating a
statistically preferred model compared with the case of no rotation.
\label{fig:balmermods}}

\end{figure*}

\begin{figure}[ht!]
   \centering
   \includegraphics[scale=.6,clip,trim=0mm 10mm 30mm 30mm,angle=0]{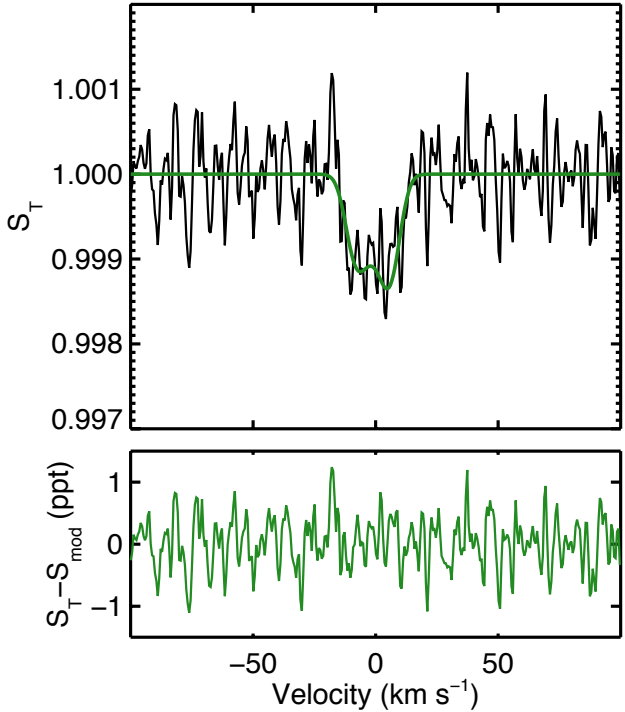}

   \figcaption{Best-fit atmospheric model to the \ion{Mg}{1} line. Although
non-zero rotation is preferred, the atmospheric parameters are highly uncertain
and the in-transit variability calls into question the spherical symmetry
assumed in the model.  \label{fig:mgimods}}

\end{figure}

The joint posterior distributions and marginalized histograms for each
parameter are shown in \autoref{fig:balmcorn} and \autoref{fig:balmcorn_norot}
for the Balmer lines and in \autoref{fig:mgicorn} for the \ion{Mg}{1} 5172.7
\AA\ line.  Note that in \autoref{fig:balmcorn} the posterior for $r$ appears
bimodal. This parameter is semi-continuous since it can only take on values in
increments of $0.01 R_*$, the spatial resolution of the grid. The best-fit
parameter values, which we take to be the median of the marginalized posterior,
and their 68\% confidence intervals are given in \autoref{tab:rotpars}.
The best-fit line profiles are shown in \autoref{fig:balmermods} and
\autoref{fig:mgimods}.

\begin{deluxetable}{lcccc}
\tablecaption{Best-fit atmospheric parameters\label{tab:rotpars}}
\tablehead{\colhead{Parameter}&\colhead{Symbol}&\colhead{Units}&\colhead{\ion{Mg}{1}}&\colhead{Balmer}}
\colnumbers
\tabletypesize{\scriptsize}
\startdata
Number density & $n$ & cm$^{-3}$ & $10.0^{+31.9}_{-7.6}$ & $836.0^{+2773.9}_{-487.8}$ \\
Radial extent & $R_\text{p}+r$ & $R_\text{p}$ & $0.6^{+0.4}_{-0.2}$ & $0.84^{+0.01}_{-0.04}$ \\
Thermal broadening & $v_\text{t}$ & km s$^{-1}$ & $4.4^{+5.9}_{-2.7}$ & $12.6^{+1.8}_{-1.7}$ \\
Rotational velocity & $v_\text{rot}$ & km s$^{-1}$ & $7.4^{+1.1}_{-1.2}$ & $8.2^{+0.6}_{-0.7}$ \\
\enddata
\end{deluxetable}




Including rotation in the atmospheric model results in a statistically
better fit for the Balmer lines: we find a difference in the Bayesian
information criterion \citep{schwarz78,liddle07} of $\Delta \text{BIC} = 76$
where the rotation model has the lower BIC. Since $\Delta \text{BIC} > 10$ is
generally considered decisive evidence in favor of the lower BIC model, we
conclude that the model value of $v_\text{rot} = 8.2^{+0.6}_{-0.7}$ km s$^{-1}$
is indeed tracing the planetary rotation. Although the \ion{Mg}{1} model fits
are less well constrained than the simultaneous fits to the Balmer lines, and
may be contaminated by non-spherically symmetric atmospheric absorption, the
value of $v_\text{rot} = 7.4^{+1.1}_{-1.2}$ km s$^{-1}$ is consistent with the
tidal-locked equatorial rotational velocity. The Balmer line value is
suggestive of mild superrotation which may be the result of higher jet
speeds at greater altitudes \citep[e.g.,][]{rauscher14}.

\subsection{Absorption time series}
\label{sec:timeseries}

We examine how the absorption in the various atomic species changes as a
function of transit time. For the Balmer lines, the absorption is strong enough
to consider the lines individually. The metal line depths are fairly shallow
($\approx 0.1-0.4$\%) but there are at least two individual lines with
significant absorption in the average transmission spectrum for each of the
metal line species.  We thus co-add the individual lines from each species to
produce an average line profile at each observation time. This increases the
S/N by a factor of $\approx \sqrt{N}$, where $N$ is the number of detected
lines for that species.  The absorption is calculated identically to the
average in-transit values. To derive uncertainties for the individual time
series points, we consider both the Poisson noise in the co-added spectrum and
the standard deviation of the out-of-transit points used in the average
out-of-transit spectrum. The larger of the two is conservatively used as the
$1\sigma$ uncertainty for each point.  In addition to the detected atomic
species, we also perform the same time series procedure for the control region
discussed in \autoref{sec:metals}. This ensures that our analysis is not
producing spurious absorption features. The individual in-transit spectra used
to calculate the time series are given in \autoref{sec:appendixa}. 

\begin{figure}[htbp!]
   \centering
   \includegraphics[scale=.82,clip,trim=60mm 55mm 5mm 0mm,angle=0]{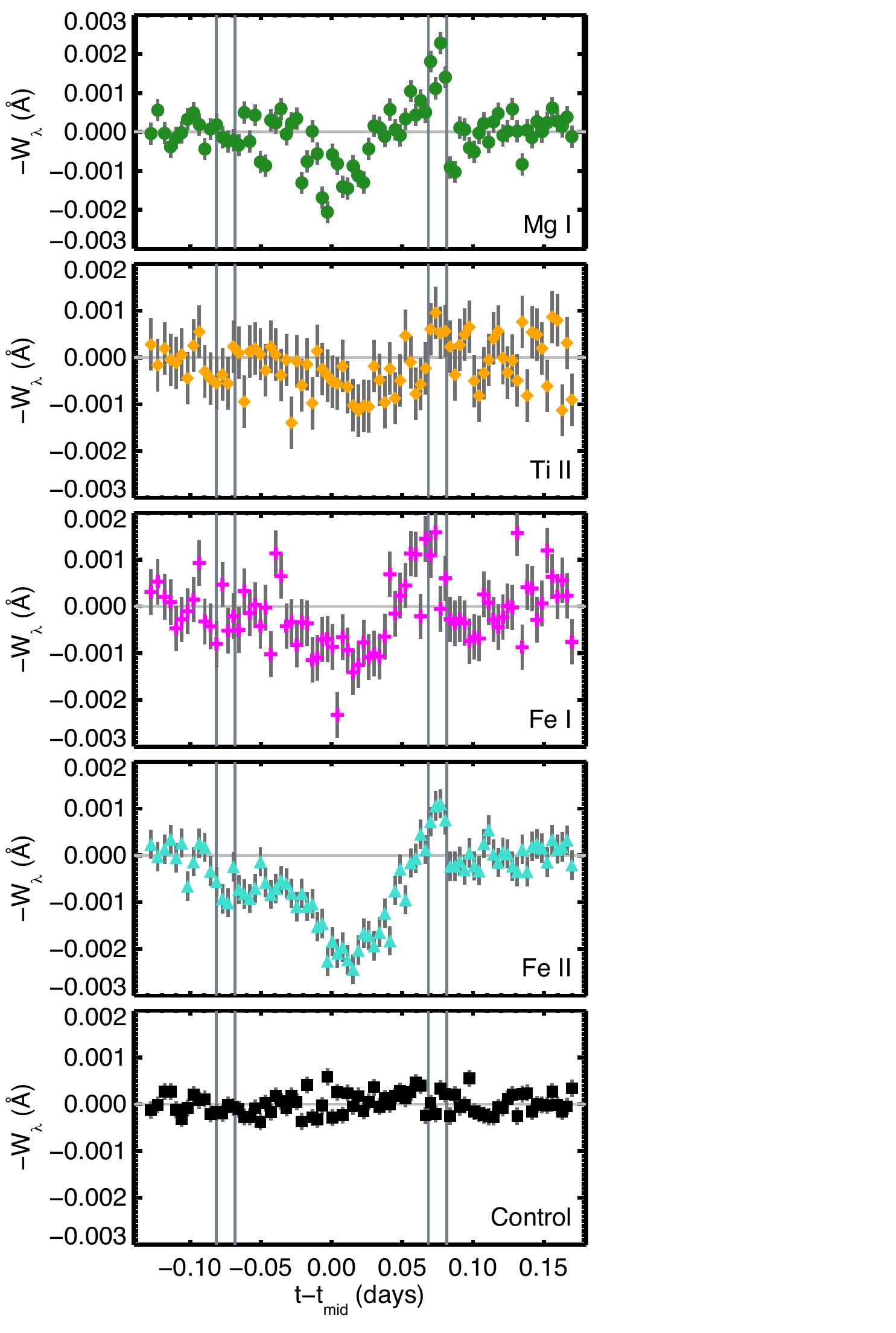}

   \figcaption{Timeseries absorption for average combined profiles of all the
detected metal lines.  The vertical gray lines mark the transit contact points
and the horizontal gray line is at $-W_\lambda = 0$.  The legend is given in
the lower right corner. There is significant structure in the light curves and
the metal absorption tends to correlate across the various lines. The positive
bump immediately before third contact is consistent across all lines and also
appears in the Balmer line time series (see \autoref{fig:balmertime}).
\label{fig:metaltime}}

\end{figure}

\begin{figure*}[h!]
   \centering
   \includegraphics[scale=.70,clip,trim=20mm 20mm 5mm 30mm,angle=0]{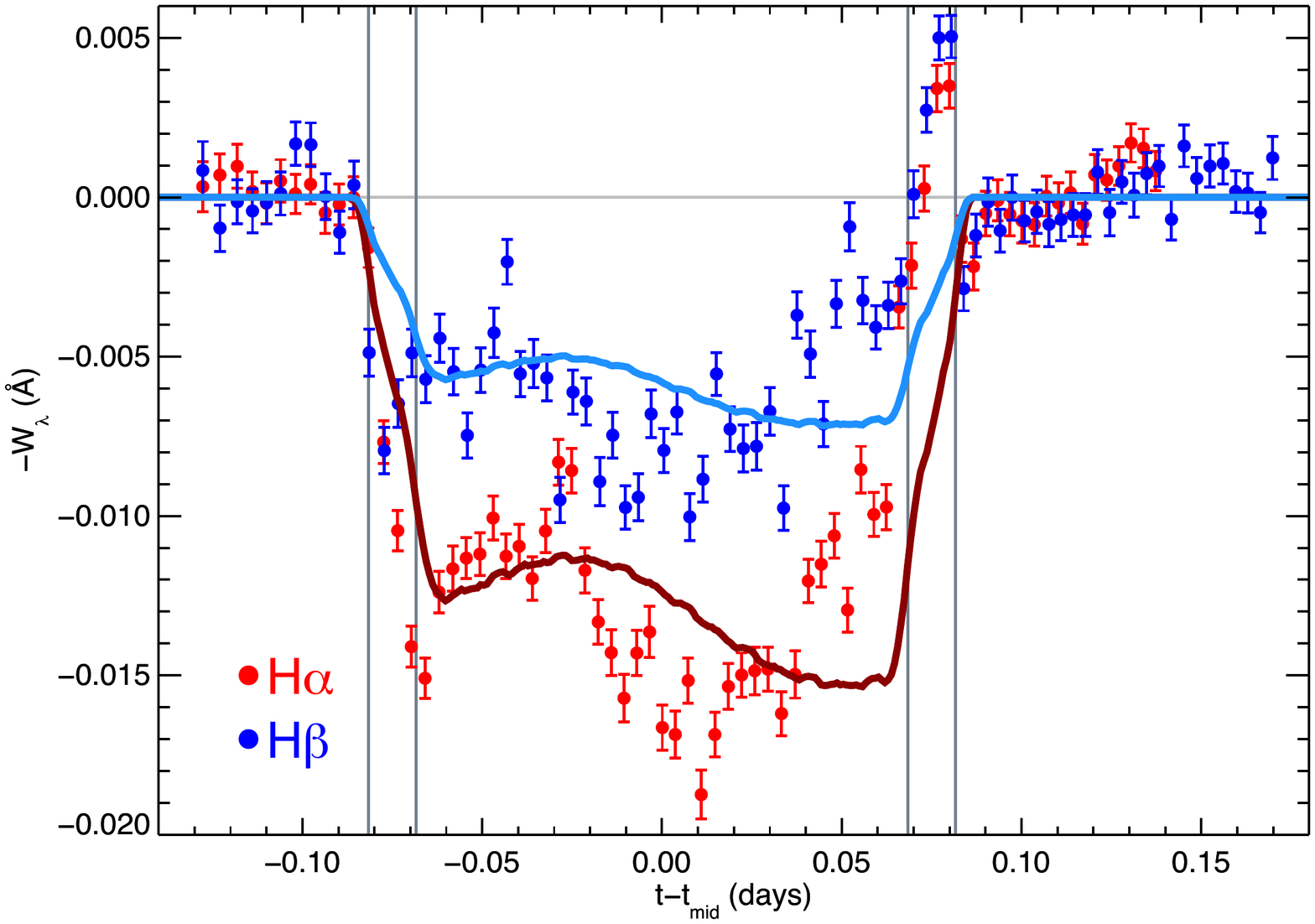}
   \figcaption{Timeseries of H$\alpha$ and H$\beta$ absorption. The favored atmosphere
model time series is overplotted with the solid dark lines. Note the deviations
from the symmetric atmosphere model and the emission feature near third contact,
which is also evident in the metal line time series in \autoref{fig:metaltime}.
\label{fig:balmertime}}
\end{figure*}

The metal time series are shown in \autoref{fig:metaltime} and the Balmer line
time series are shown in \autoref{fig:balmertime}. In both plots the transit
contact points are marked with vertical gray lines. The time series absorption
calculated with the favored Balmer line model from \autoref{sec:atmomods} is
shown with solid lines in \autoref{fig:balmertime}. In \autoref{fig:metaltime}
the time series for the control region is shown in black in the bottom panel.

The timeseries show significant structure: most of the metal lines show
initially weak absorption, increasing rather sharply towards mid-transit and
then abruptly decreasing before third contact. The Balmer time series show very
similar structure and are not well represented by the spherically symmetric
atmosphere models from \autoref{sec:atmomods} (solid lines in
\autoref{fig:balmertime}). 

Strikingly, all of the lines show a marked \textit{emission}-like feature
immediately before third contact which disappears immediately after fourth
contact. Since $S_\text{T}$ is a relative measurement of spectral lines
observed at different times, emission features could also be the lack of
absorption during one exposure compared with another. However, the stability of
the out-of-transit points suggests that the emission bump is indeed a filling
in of the nominal line core; a relative decrease in absorption would likely
show a continuing trend during the out-of-transit exposures since these 
points would still be in absorption relative to the emission peak.

\section{DISCUSSION}
\label{sec:discussion}

Here we discuss the atmospheric model line profiles and possible explanations
for the origin of the in-transit absorption variability.

\subsection{Atmospheric rotation}
\label{sec:atmorot}

We showed that the Balmer line and \ion{Mg}{1} average in-transit transmission
spectra are well-fit with rotational velocities of $7.4^{+1.1}_{-1.2}$ km
s$^{-1}$ and $8.2^{+0.6}_{-0.7}$ km s$^{-1}$, respectively. The \ion{Mg}{1}
value of $v_\text{rot}$ is consistent with KELT-9 b being tidally locked to its
host. The Balmer line value, however, suggests that KELT-9 b is mildly
supperotating.  We do not take into account the effects of jets or
day-to-nightside winds in KELT-9 b's thermosphere, which should produce a bulk
blue-shift in the transmission spectrum. These features naturally occur in hot
planet atmospheres \citep[e.g.,][]{showman08,rauscher10} and have been
marginally detected in HD 209458 b \citep{snellen10} and HD 189733 b
\citep{louden15}. With the potential exception of H$\beta$, no significant net
blue-shifts are seen in our spectra (see \autoref{tab:abslines}). However, as
discussed in \autoref{sec:atmomods}, velocities associated with the in-transit
variability may mask the nominal atmospheric wind velocities. 

The difference in the model $v_\text{rot}$ values may also be due to the
different pressure regions probed by the Balmer lines compared with the
\ion{Mg}{1} triplet. \citet{kempton12} showed that wind speeds of up to $\approx
15$ km s$^{-1}$ can form in atmospheres with no magnetic drag, at pressure
consistent with the formation of the Balmer lines. Rotation combined with jet
speeds of $\approx 2-3$ km s$^{-1}$ could be enough to explain the large
$v_\text{rot}$ value for the Balmer lines. Future non-variable transits should
provide cleaner opportunities to measure the normal atmospheric motions of
KELT-9 b. Modeling efforts aimed at the optical \ion{Mg}{1} triplet would
also be helpful in breaking the degeneracy between the density and radial
extent of the atmosphere.

\subsection{Estimating the mass loss rate of KELT-9 b}
\label{sec:massloss}

Thermal atmospheric escape is predicted to be a ubiquitous phenomenon for hot
gas giants \citep[e.g.,][]{murray09,salz16} and has been calculated in a handful
of systems using models of the high-velocity wings of Lyman-$\alpha$
\citep{vidal03,desetangs10,ehrenreich12,kulow14,ehrenreich15}. \citet{gaudi17}
estimated a mass loss rate of $\dot{M} \approx 10^{10} - 10^{13}$ g s$^{-1}$.
where the large range is due to uncertainty concerning KELT-9's emitted
non-thermal radiation and stellar activity level. \citet{yan18} were able
to use an analytic atmospheric model to derive a density based on their
observed H$\alpha$ profile and estimate a mass loss rate of $\dot{M} \approx
10^{12}$ g s$^{-1}$.

We can use the densities derived from our atmospheric models to calculate an
approximate mass loss rate for KELT-9 b. For the case of \ion{Mg}{1}, the
number density of the triplet ground state is $n=10.0$ cm$^{-3}$. We assume
that the material is moving at $10$ km s$^{-1}$ as it passes through the model
value $r=1.6 R_\text{p}$.  If we assume statistical equilibrium for the
magnesium atoms and a thermospheric temperature of $T=10,000$ K, the ratio of
Mg I triplet states to the Mg I ground state is $\approx 0.2$. If we further
assume that half of all the magnesium atoms are ionized, we arrive at a total
mass loss rate for magnesium of $\dot{M}_{MgI} \approx 3.6\times10^7$ g
s$^{-1}$. Finally, assuming a solar magnesium abundance for KELT-9 b
\citep{asplund09}, we calculate an approximate total mass loss rate of $\dot{M}
\approx 1\times10^{12}$ g s$^{-1}$.  Performing a similar exercise with the
Balmer line atmospheric parameters, and assuming $n_2/n_1 = 10^{-6}$, we
calculate $\dot{M} \approx 3\times10^{12}$ g s$^{-1}$, similar to the \ion{Mg}{1} value.
 
Mass loss calculations of this sort are highly uncertain due to assumptions
concerning the level populations, outflow velocity, planetary abundances,
and ionization state. However, it is encouraging that both of our estimates,
which are based on the density and radial extent derived from atmospheric
modeling, are within the range expected from energy deposition calculations
and agree with the value from \citet{yan18}. Atmospheric models of
ground-state \ion{Mg}{1} absorption coupled with simultaneous observations
of the excited state \ion{Mg}{1} optical triplet could provide strict
constraints on a planet's mass loss rate. 

\subsection{In-transit absorption variability and the emission feature}
\label{sec:intranvar}

Both the metal line and Balmer line time series display interesting
sub-structure. This is most visible in the Balmer line absorption in
\autoref{fig:balmertime} but can also be clearly seen in the \ion{Fe}{2} panel
in \autoref{fig:metaltime}: the absorption is initially fairly weak and, in the
case of the Balmer lines, appears to correspond to the symmetric atmosphere
model and then begins to increase near mid-transit before sharply decreasing
around $t-t_\text{mid} \approx 0.05$ days, i.e., the emission feature mentioned
in \autoref{sec:timeseries}. 

We can gain additional information by looking at the velocities of the
transmission line profiles. Due to the low signal-to-noise of most of the metal
lines, and some of the H$\beta$ lines, we focus on the the individual H$\alpha$
and \ion{Fe}{2} profiles (see \autoref{fig:halphaall} and
\autoref{fig:feiiall}). The velocities, or what we will call $V_0$, are
calculated as a flux-weighted average of the planetary-frame
transmission spectrum between $\pm 50$ km s$^{-1}$. The uncertainties are the
standard deviation of the velocities which contribute 68\% of the transmission
spectrum absorption or emission across the $\pm 50$ km s$^{-1}$ velocity range.
Thus broader profiles have larger uncertainties compared with narrow profiles.

\begin{figure}[h!]
   \centering
   \includegraphics[scale=.4,clip,trim=30mm 20mm 5mm 30mm,angle=0]{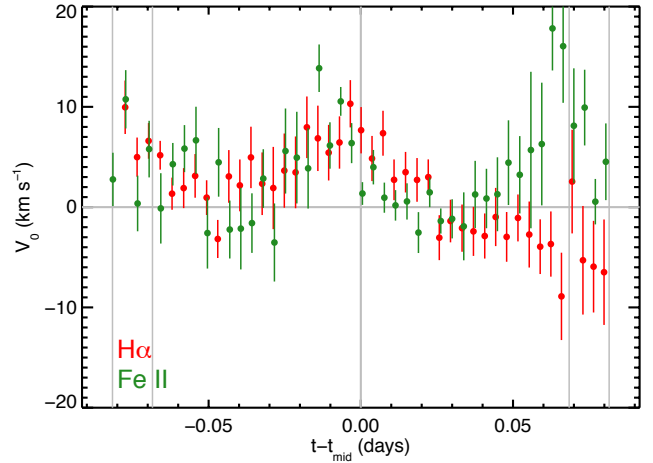}

   \figcaption{Flux-weighted velocity measurements $V_0$ for the individual
H$\alpha$ and \ion{Fe}{2} transmission spectra in \autoref{fig:halphaall} and
\autoref{fig:feiiall}.  The vertical gray lines show the transit contact points
and $t-t_\text{mid} = 0.0$ days.  The lines become red-shifted when the
absorption increases around $t-t_\text{mid} \approx -0.03$ days and then
gradually change to blue-shifted. The \ion{Fe}{2} lines become sharply
red-shifted when the emission features start to become prominent.
\label{fig:linevels}}

\end{figure}

The in-transit H$\alpha$ and \ion{Fe}{2} velocities are shown in
\autoref{fig:linevels}. Initially, $V_0$ is close to 0 km s$^{-1}$ for
both lines, which is expected in the planet's rest frame in the absence of
other asymmetric velocity components. The velocities in both lines then become
strongly red-shifted near $t-t_\text{mid} \approx -0.03$ days, similar to when
the in-transit absorption increases. This occurs in both H$\alpha$ and
\ion{Fe}{2}, suggesting that both atomic populations are tracing similar
regions in the atmosphere. Near $t-t_\text{mid} \approx 0.05$ days, when the
emission features appear, the \ion{Fe}{2} velocities become strongly
red-shifted, due to the emission peaks in the spectra, and H$\alpha$ continues
to be mildly blue-shifted, although most of the measurements are consistent
with $V_0 \approx 0.0$ km s$^{-1}$. The velocity variations cannot be
caused by uncertainty in the planet's orbital velocity: changing the in-transit
line-of-sight velocity of the planet would leave a linear residual in the
transmission profile velocities, not the non-uniform variations that are
observed.

\begin{figure}[h!]
   \centering
   \includegraphics[scale=.55,clip,trim=40mm 20mm 60mm 70mm,angle=0]{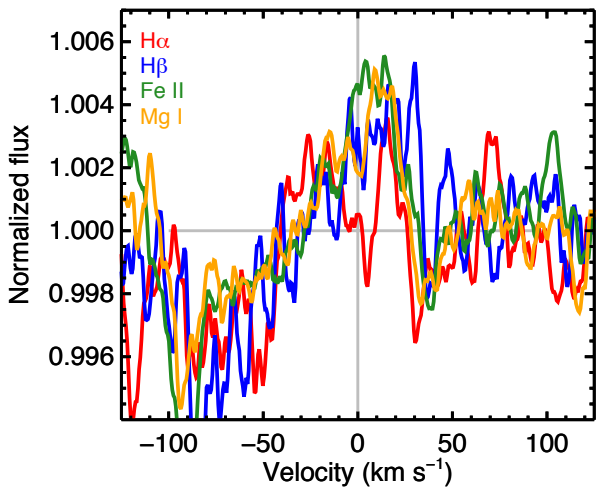}

   \figcaption{Average transmission spectra for H$\alpha$, H$\beta$,
\ion{Mg}{1}, and \ion{Fe}{2} in the planetary rest frame for times between
third and fourth contact where the emission peaks in \autoref{fig:metaltime}
and \autoref{fig:balmertime}. The spectra are smoothed by 10 pixels for clarity
and the \ion{Fe}{2} and \ion{Mg}{1} spectra have been scaled to make them
comparable in strength to the Balmer line spectra. The line profiles show
P-Cygni-like shapes: red-shifted emission and blue-shifted absorption,
indicative of outflowing material.  \label{fig:emission}}

\end{figure}

The metal and Balmer line timeseries exhibit a significant emission-like
feature near third contact. Shown in \autoref{fig:emission} are the average
$S_\text{T}$ spectra for the Balmer lines, \ion{Mg}{1}, and \ion{Fe}{2} for
times between third and fourth contact. There is remarkable consistency among
the line profiles: blue-shifted absorption extending to $\approx 100$ km
s$^{-1}$ and an emission peak that is red-shifted by $\approx 10$ km s$^{-1}$.
The exception is H$\alpha$, which shows a weaker emission profile: two of the
four spectra between third and fourth contact are in absorption, decreasing the
emission strength. This is likely due to optical depth effects which prevent
the absorption line core from being entirely filled in. We reiterate
that the large blue-shifted velocities cannot be the result of uncertainty in
the planet's orbital velocity, which is at most $\approx 6-8$ km s$^{-1}$
\citep{gaudi17,yan18}.  

The profile shapes in \autoref{fig:emission} are reminiscent of P-Cygni
features, the result of strong stellar winds around luminous variable stars
\citep[e.g.,][]{martins15,gillet16} or accreting pre-main sequence stars
\citep[e.g.,][]{edwards03,cauley14,cauley15a}. The blue-shifted absorption
suggests a wind-like geometry where material is being accelerated away from the
planet to velocities of $\approx 50-100$ km s$^{-1}$ in the direction of the
observer.  The P-Cygni profile disappears immediately after transit,
constraining the geometry of the material to the immediate vicinity of the
planet, i.e., a trailing tail of material would produce a post-transit
absorption signature relative to the pre-transit observations.

Thermal expansion from the planet's atmosphere alone cannot account for such
large blue-shifted velocities \citep[e.g.,][]{trammell14,salz16}. Radiation
pressure from the star is the only plausible mechanism capable of accelerating
the planetary wind material to high velocities. Radiation pressure has been
shown to strongly affect the geometry of planetary mass flows in other systems
\citep[e.g.,][]{bourrier13,bourrier16,spake18}.  

Neglecting other forces, the ratio $\beta = F_\text{rad}/F_\text{g}^\star$, where
$F_\text{rad}$ is the force due to radiation pressure on an atom and
$F_\text{g}^\star$ is the stellar gravitational force, determines whether a
particle at rest is pushed away ($\beta > 1$) or falls towards the star ($\beta
< 1$). The value of $\beta$ for a particular atomic species is independent of
the distance from the star since both $F_\text{rad}$ and $F_\text{g}^\star$ go like
$r^{-2}$.  Values of $\beta$ for some the atomic species of interest here have
been calculated for systems similar to KELT-9: \citet{beust01} investigated the
observational signatures of falling evaporating bodies around Herbig Ae/Be
stars and calculated $\beta$ for \ion{Fe}{1}, \ion{Fe}{2}, and \ion{H}{1} in
the case of $T_\text{eff} = 10500$ K, $M_\star = 2.4 M_\odot$, $R_\star = 1.75
R_\odot$, and log$g=4.0$. They find $\beta_\text{FeII} = 46.4$,
$\beta_\text{FeI} = 90.9$, and $\beta_\text{HI} = 0.43$.  Although
$\beta_\text{HI}<1$ suggests that radiation pressure is not sufficient to
accelerate hydrogen atoms, the hydrogen escaping from KELT-9 b is, in addition
to the thermal expansion velocity away from the planet, in orbit and thus even
moderate additional radial forces are capable of pushing hydrogen to larger
velocities. 

If we adopt the \citet{beust01} $\beta$ values for the KELT-9 system, is
radiation pressure capable of producing the $\approx 50-100$ km s$^{-1}$
blue-shifts seen in the line profiles? We assume that the only forces acting on
the planetary material are the planetary gravity and radiation pressure. We
ignore the stellar gravitational potential since $F_\text{g}^\text{pl} \approx
1.6 \times F_\text{g}^\star$ and the felt centripetal acceleration on the
particles, in the absence of perturbations, balances the stellar gravity. For
KELT-9 b, $a_\text{g}^\text{pl} = 0.020$ km s$^{-2} = 1.59 a_\text{g}^\star$.
For \ion{Fe}{2} particles moving radially away along the star-planet line, the
net acceleration is then $a_\text{net} \approx 0.56$ km s$^{-2}$ away from the
planet. Under constant acceleration, an \ion{Fe}{2} atom could reach velocities
of $\approx 100$ km s$^{-1}$ in $\approx 3$ minutes. 

The above idealized calculation ignores self-shielding of the atmosphere, which
reduces $\beta$ and thus the acceleration due to radiation pressure. It also
ignores any initial velocity given to the particles by the planetary wind.
Nonetheless, the order-of-magnitude estimates suggest that radiation is capable
of producing the blue-shifted velocities seen in \autoref{fig:emission}. Note
that the blue-shifted velocities in the line profiles in \autoref{fig:emission}
are outside of the velocity range used to calculate $V_0$ in
\autoref{fig:linevels}. Thus the values of $V_0$ between third and fourth
contact are tracing the mildly red-shifted emission lines for \ion{Fe}{2}. If
the material is strongly accelerated radially away from the star-planet line,
this could produce the geometry necessary to explain the disappearance of the
absorption immediately after transit: a narrow cometary tail will form
behind the planet with little azimuthal extent. This is similar to what is
predicted for WASP-107 b based on the detection of \ion{He}{1} 10830 \AA\
\citep{oklopcic18,spake18}.

The in-transit absorption and velocity variations are indicative of dynamic
circumplanetary material or time variability in the density and structure of
the extended atmosphere. Another possibility is that variations in stellar
activity or the transiting of quiescent stellar active regions are affecting
the transmission spectra. However, KELT-9 is an A0V/B9V star and it is unlikely
that it has the active region coverage fraction or intrinsic active region
contrast necessary to produce such changes in the Balmer lines via transits of
bright active regions \citep{cauley18}.  An increase in stellar brightness
would cause a decrease in the transmission spectrum strength. Thus the
in-transit variations are probably intrinsic to changes in the planet's
extended atmosphere.

We propose the following scenario to explain both the absorption increase near
$t-t_\text{mid} \approx -0.03$ and the large blue-shifted velocities near the
end of the transit: a stellar flare caused an inflation of the planet's
extended atmosphere and a spike in the mass loss rate, creating the enhanced
absorption around $t-t_\text{mid} \approx -0.03$ days. The expanding material
escaped and was accelerated away from the planet, which produces the P-Cygni
line profiles. 

Although the presence of flares on A-stars is debated
\citep[e.g.,][]{balona15,pedersen17}, we now know of spots
\citep{bohm15,petit17} and small-scale magnetic fields on normal A0V stars
\citep[e.g., Vega;][]{lignieres09,petit10}. Thus it is not implausible that a
small flare event, perhaps triggered by an interaction between the stellar and
planetary magnetic fields \citep[e.g.,][]{lanza09,strugarek14}, is responsible
for the observed phenomena.  Future spectroscopic transit observations with
simultaneous photometry will help determine the origin of these features and
quantify how common they are for this system.

\section{SUMMARY AND CONCLUSIONS}
\label{sec:conclusions}

We observed the transit of the ultra-hot Jupiter KELT-9 b using the
high-resolution spectrograph PEPSI on the LBT. In addition to confirming the
detections of \ion{Ti}{2}, \ion{Fe}{1}, \ion{Fe}{2}, and H$\alpha$ in the
atmosphere of KELT-9 b, we report the precense of H$\beta$ and the first
detection of the optical \ion{Mg}{1} triplet in an exoplanet atmosphere.
Magnesium is an important coolant in exoplanet atmospheres \citep{huang17} and
may be a useful diagnostic of evaporation \citep{bourrier15}. Mass loss
rates calculated from the Balmer line and \ion{Mg}{1} atmospheric
parameters agree well with previous estimates from \citet{gaudi17} and
\citet{yan18}. 

We reproduced the extended hydrogen envelope absorption with rotating
atmosphere models and found that rotation velocities of $\approx 8$ km s$^{-1}$
are strongly preferred to the case of no rotation. This suggests that
the Balmer lines trace a region of the atmosphere that is mildly superrotating
compared with the tidally locked rotation rate. Models of the \ion{Mg}{1}
absorption suggest a rotation rate consistent with the planet being tidally
locked, although the \ion{Mg}{1} atmospheric model parameters are much more
uncertain. The in-transit absorption variability likely complicates the line
\ion{Mg}{1} profile shape, with possible contributions to the profile velocity
information from mass flows associated with evaporating material, making the
precise interpretation of the rotational velocity difficult. Observations of a
more uniform transit will allow for better constraints on the rotational
information contained in the \ion{Mg}{1} line profiles.

The time-resolved transit absorption shows significant variations in all of the
detected atomic species, including the hydrogen Balmer lines, and suggests that
the transiting planetary material is spatially non-uniform or varying in time.
An emission feature at the end of the transit is suggestive of mass loss from
the planet and shows spectroscopic features similar to strong stellar winds,
evidence that the evaporating material is being rapidly accelerated away from
the planet, likely by radiation pressure from the star. We suggest that the
sudden in-transit absorption variability, and subsequent appearance of P-Cygni
profiles in the transmission spectra, are evidence of a flaring event that
resulted in enhanced mass loss and thus a larger extended atmosphere. More
spectroscopic transits, combined with simultaneous photometry, are needed to
determine the frequency of such behavior and help understand the physical cause
of the in-transit variability.

{\bf Acknowledgments:} P. W. C. and E. L. S. are grateful for support from NASA
Origins of the Solar System grant No. NNX13AH79G (PI: E.L.S.). It is also our
pleasure to thank the German Federal Ministry (BMBF) for the year-long support
for LBT/PEPSI through their Verbundforschung with grants 05AL2BA1/3 and
05A08BAC. S. R. acknowledges support from the National Science Foundation
through Astronomy and Astrophysics Research Grant AST-1313268. A. G. J. wishes
to acknowledge support through NASA Exoplanet Research Program grant
14-XRP14-2-0090 to the University of Nebraska at Kearney. This work has made
use of NASA's Astrophysics Data System.

\software{\texttt{Molecfit}, \citet{kausch15}, \texttt{Spectroscopy Made Easy
(SME)}, \citet{valenti96,piskunov17}}

\clearpage

\appendix
\section{Individual transmission spectra}
\label{sec:appendixa}

In this appendix we show the in-transit spectra for each of the lines presented
in \autoref{sec:results}. These line profiles are used to calculate the values
shown in \autoref{fig:metaltime} and \autoref{fig:balmertime}. 

\begin{figure*}[h!]
   \centering
   \includegraphics[scale=.85,clip,trim=5mm 10mm 0mm 0mm,angle=0]{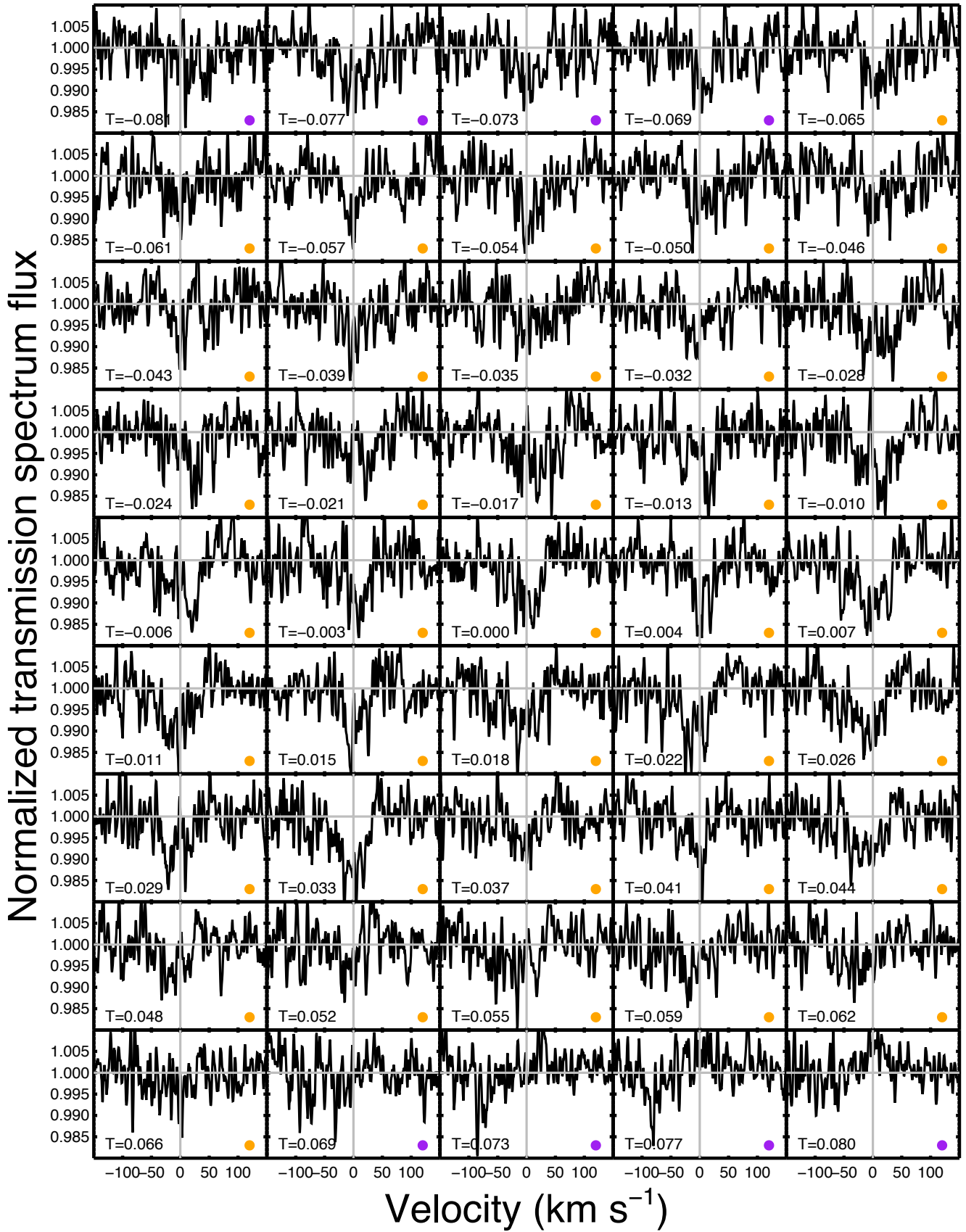}

   \figcaption{Individual in-transit transmission spectra of H$\beta$. The
transit time in days from mid-transit is given in the lower left of each panel.
Panels marked with a purple dot signify transit times between 1st and 2nd or
3rd and 4th contact; orange dots mark spectra taken between 2nd and 3rd
contact, which are the spectra used to create the average in-transit profiles
shown in \autoref{fig:tspecs_all}. All spectra are smoothed with a 5-pixel
boxcar kernel for clarity.  \label{fig:hbetaall}}

\end{figure*}

\begin{figure*}[htbp]
   \centering
   \includegraphics[scale=.85,clip,trim=5mm 10mm 0mm 0mm,angle=0]{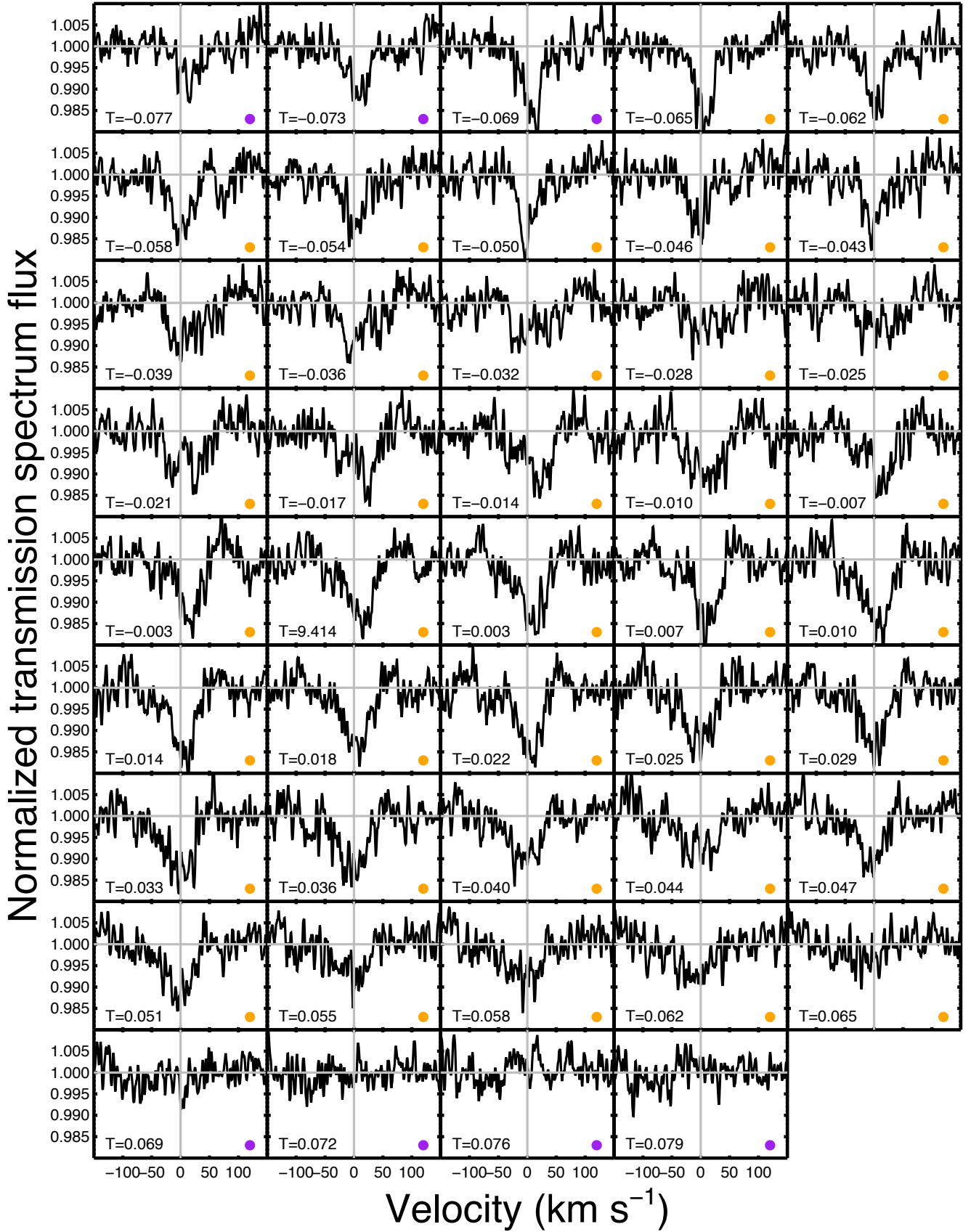}
   \figcaption{Individual in-transit transmission spectra of H$\alpha$. The format is the same
   as \autoref{fig:hbetaall}.
\label{fig:halphaall}}
\end{figure*}

\begin{figure*}[htbp]
   \centering
   \includegraphics[scale=.85,clip,trim=5mm 10mm 0mm 0mm,angle=0]{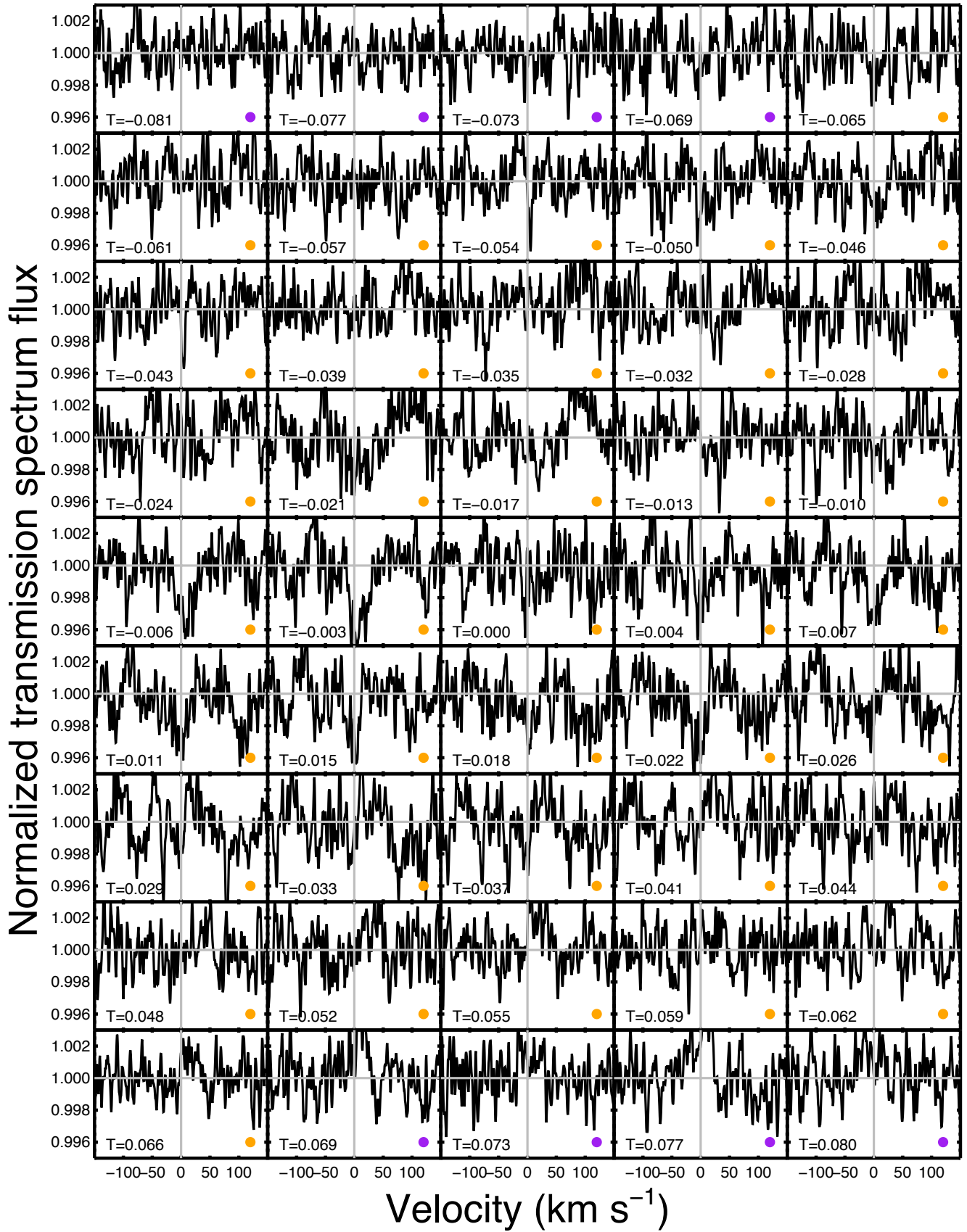}
   \figcaption{Individual in-transit transmission spectra of the combined \ion{Mg}{1} line
profile. The format is the same as \autoref{fig:hbetaall}. Note the \ion{Fe}{2} 5169.0 \AA\
line near $+100$ km s$^{-1}$.
\label{fig:mgiall}}
\end{figure*}

\begin{figure*}[htbp]
   \centering
   \includegraphics[scale=.85,clip,trim=5mm 10mm 0mm 0mm,angle=0]{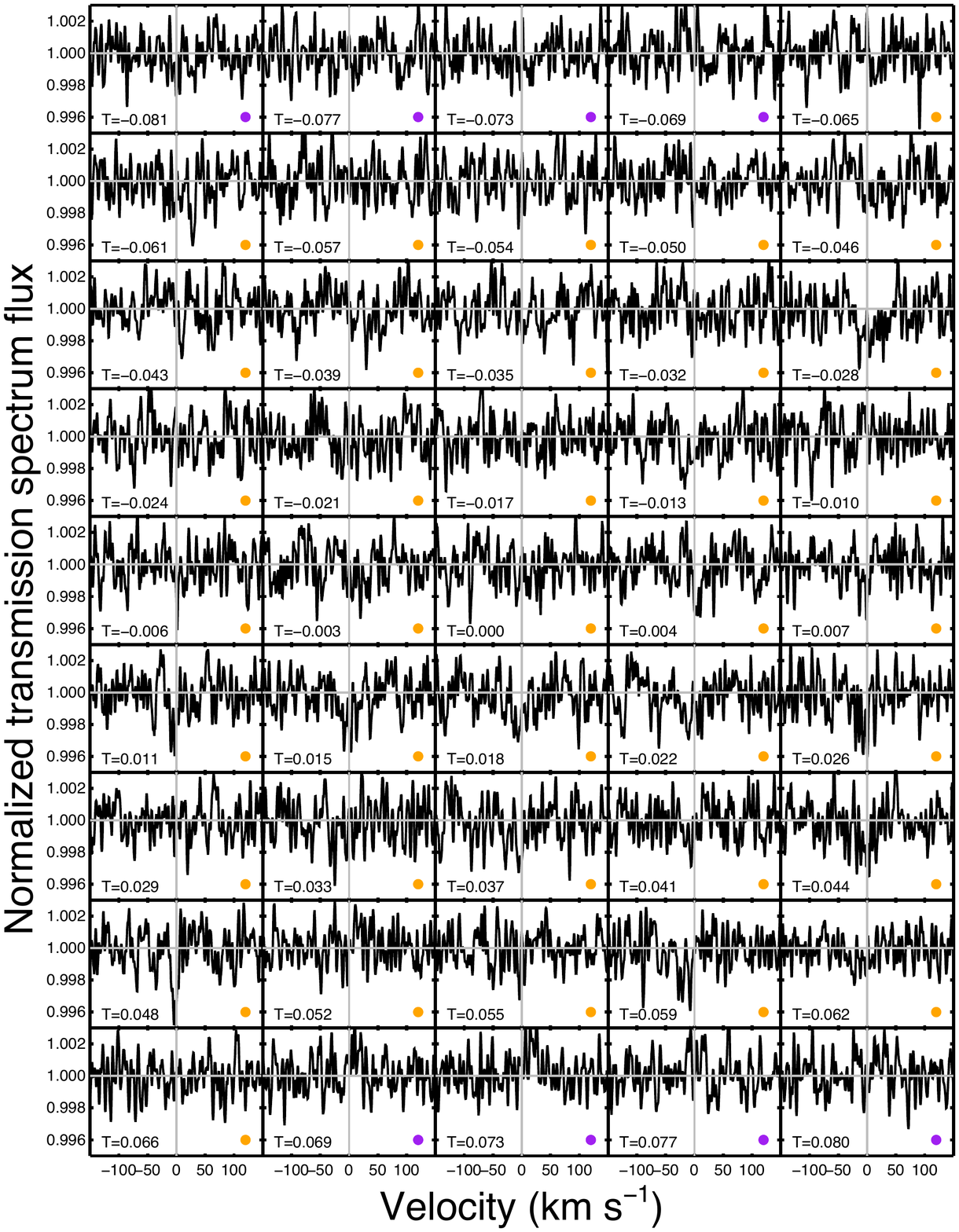}
   \figcaption{Individual in-transit transmission spectra of the combined \ion{Ti}{2} line
profile. The format is the same as \autoref{fig:hbetaall}.
\label{fig:tiiall}}
\end{figure*}

\begin{figure*}[htbp]
   \centering
   \includegraphics[scale=.85,clip,trim=5mm 10mm 0mm 0mm,angle=0]{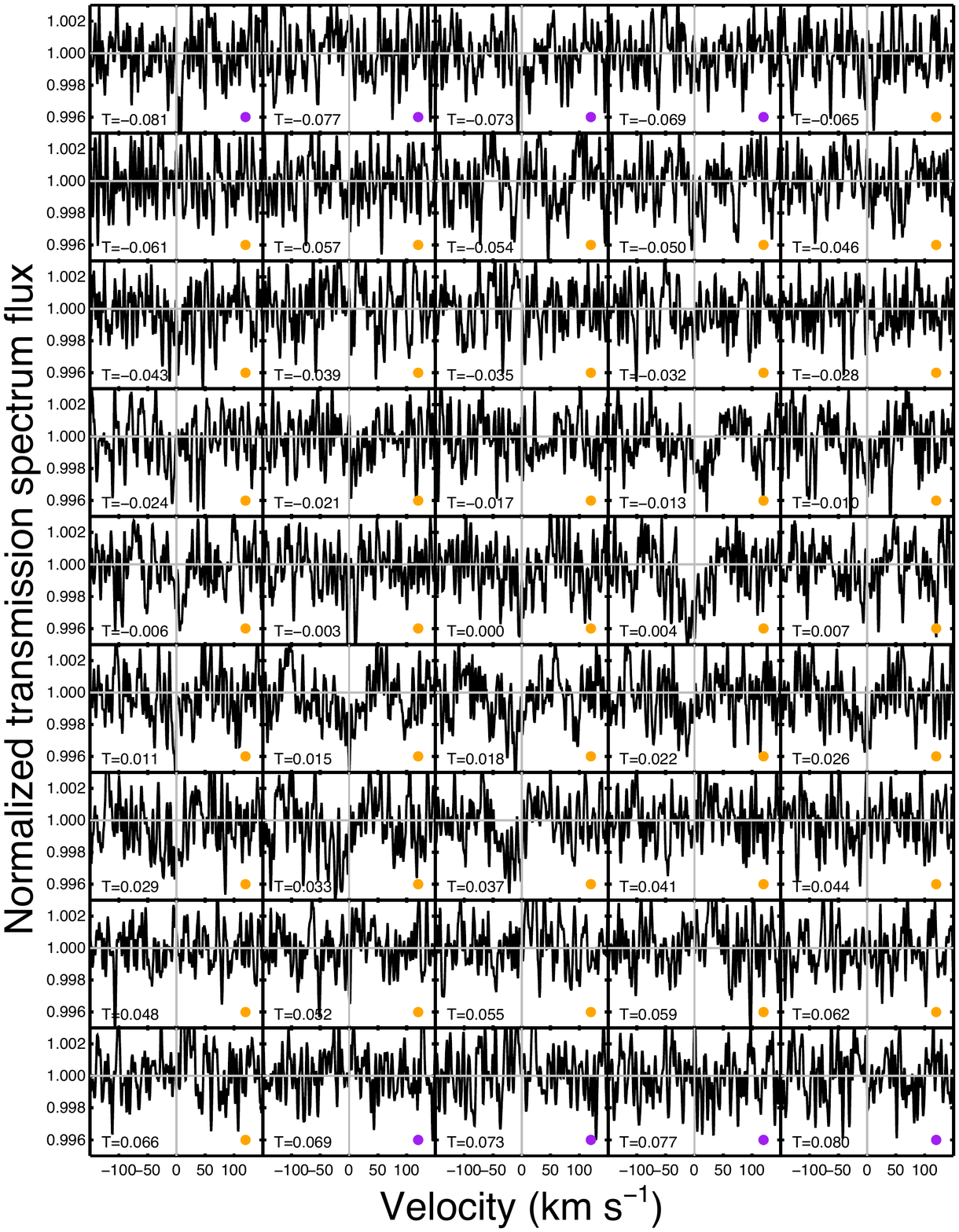}
   \figcaption{Individual in-transit transmission spectra of the combined \ion{Fe}{1} line
profile. The format is the same as \autoref{fig:hbetaall}.
\label{fig:feiall}}
\end{figure*}

\begin{figure*}[htbp]
   \centering
   \includegraphics[scale=.85,clip,trim=5mm 10mm 0mm 0mm,angle=0]{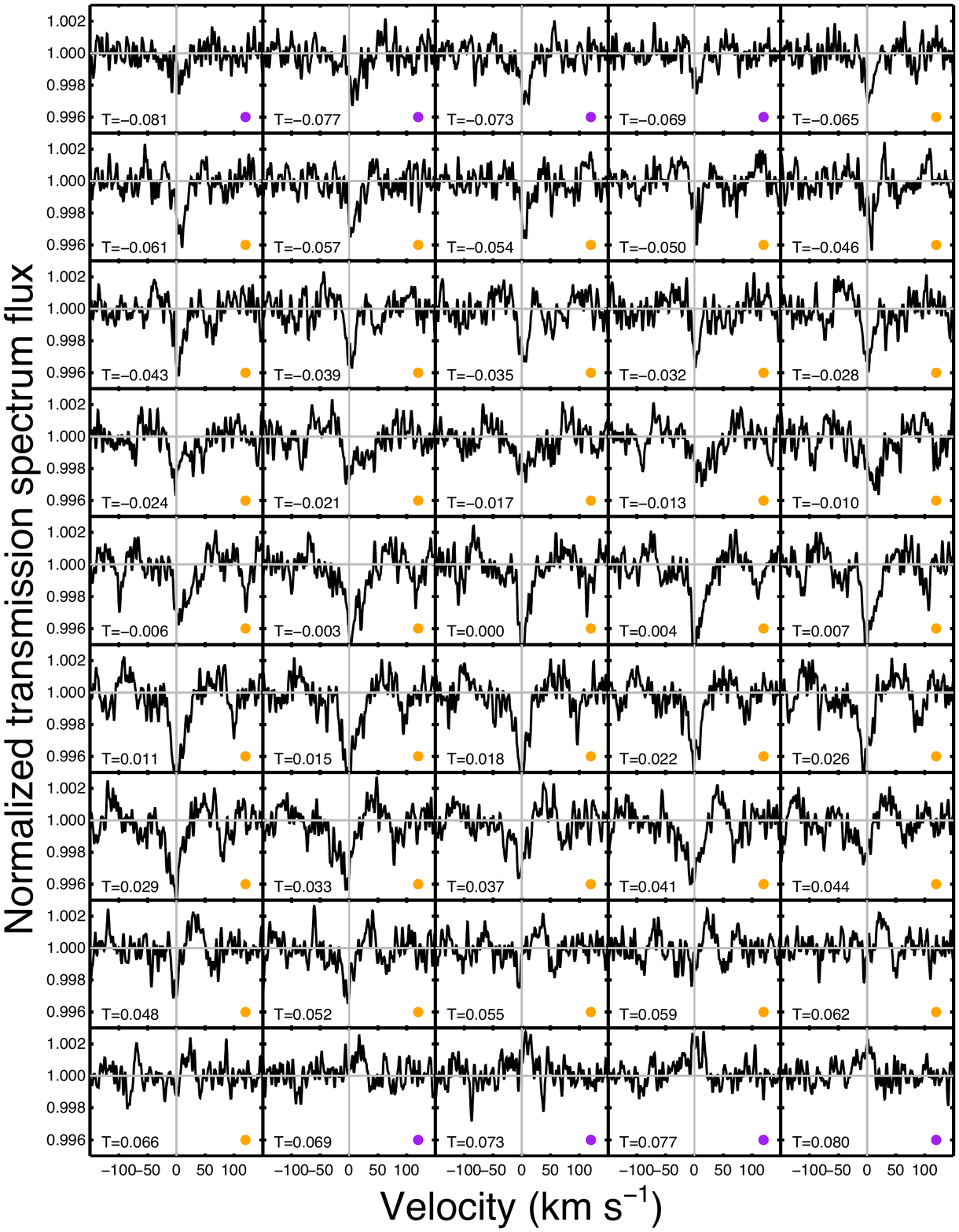}
   \figcaption{Individual in-transit transmission spectra of the combined \ion{Fe}{2} line
profile. The format is the same as \autoref{fig:hbetaall}.
\label{fig:feiiall}}
\end{figure*}

\end{document}